\documentclass[
onecolumn,
showkeys,
longbibliography,
nofootinbib,
superscriptaddress,
aps,
notitlepage
]{revtex4-1}
\usepackage[margin=1in]{geometry}

\usepackage{url}
\usepackage[utf8]{inputenc}
\usepackage{amsmath}
\usepackage{amssymb}
\usepackage{pifont}
\usepackage{bm}
\usepackage{enumitem}
\usepackage[acronym]{glossaries}
\usepackage{graphicx}
\usepackage[colorlinks,unicode]{hyperref}
\usepackage[capitalise]{cleveref}  % must come after hyperref
\usepackage{mathtools}
\usepackage{multirow}
\usepackage{physics}
\usepackage{siunitx}

\usepackage[dvipsnames,table,xcdraw]{xcolor}
\usepackage{minted}
\usepackage[normalem]{ulem}
\newacronym{dm}{DM}{dark matter}
\newacronym{ce}{CE}{Cosmic Explorer}
\newacronym{et}{ET}{Einstein Telescope}
\newacronym{snr}{SNR}{signal-to-noise ratio}

\usepackage{hyperref}
\newcommand{\gammacr}{\gamma_{\mathrm {cr}}}

\newcommand{\be}{\begin{equation}}
\newcommand{\ee}{\end{equation}}

\newcommand{\Meq}{M_{\mathrm{eq}}}

\newcommand{\aeq}{a_{\mathrm{eq}}}

\newcommand{\C}{g}

\makeatletter
\renewcommand\onecolumngrid{%
\do@columngrid{one}{\@ne}%
\def\set@footnotewidth{\onecolumngrid}%
\def\footnoterule{\kern-6pt\hrule width 1.5in\kern6pt}%
}
\makeatother

\newcommand\myshade{80}
\colorlet{mylinkcolor}{ForestGreen}
\colorlet{mycitecolor}{Red}
\colorlet{myurlcolor}{violet}

\hypersetup{
  linkcolor  = mylinkcolor!\myshade!black,
  citecolor  = mycitecolor!\myshade!black,
  urlcolor   = myurlcolor!\myshade!black,
  colorlinks = true
}

\definecolor{codebg}{rgb}{0.95,0.95,0.92}

\definecolor{terminalbg}{RGB}{30,30,30}
\definecolor{terminalfg}{RGB}{230,230,230}

\DeclareSIUnit\solarmass{\ensuremath{\mathrm{M}_\odot}}
\DeclareSIUnit\parsec{pc}
\DeclareSIUnit\gigaparsec{Gpc}
\DeclareSIUnit\year{yr}

\newcommand{\milan}{Dipartimento di Fisica ``G. Occhialini'', 
Universit\`a degli Studi di Milano-Bicocca, Piazza della Scienza 3, 20126 Milano, Italy}
\newcommand{\infn}{INFN, Sezione di Milano-Bicocca, 
Piazza della Scienza 3, 20126 Milano, Italy}

\begin{document}

%---- Title
\title{\protect\path{pbhstat}: A Python package for calculating the primordial black hole abundance}

\author{Philippa S. Cole}
\email{philippa.cole@unimib.it}
\affiliation{\milan}
\affiliation{\infn}

\author{Jacopo Fumagalli}
\email{jfumagalli@fqa.ub.edu}
\affiliation{Departement de F\'isica Qu\`antica i Astrofisica and Institut de Ci\`encies del Cosmos (ICC), Universitat de Barcelona, Mart\'i i Franqu\`es 1, 08028 Barcelona, Spain}

\begin{abstract}
We present \texttt{pbhstat}, a publicly available Python package designed to compute the mass function and total abundance of primordial black holes (PBHs) from a given primordial power spectrum. The package offers a modular framework using multiple statistical approaches, including Press–Schechter theory, peaks theory, and formalisms based on the non-linear compaction function. Currently, the implementation is limited to scenarios with nearly Gaussian initial conditions.
\end{abstract}

\maketitle

\section{Introduction}

The possibility that dark matter consists of primordial black holes (PBHs) presents an intriguing avenue from both theoretical and observational perspectives \cite{Zeldovich:1967lct, Hawking:1971ei, Carr:1974nx, Page:1976df, MacGibbon:1991tj, Barrau:2003nj, Sasaki:2016jop, Carr:2020gox, Green:2020jor,Khlopov_2010}. We focus on the case in which PBHs are seeded by large-amplitude inflationary perturbations \cite{Ivanov:1994pa,Carr:1994ar,Garcia-Bellido:1996mdl}. In this context, accurately determining the PBH abundance remains highly sensitive to the interplay between various contributing factors (see \cite{yoo2024basicsprimordialblackhole,Carr:2021bzv,Escriva:2022duf} for reviews on the subject). Central to this is the choice of collapse statistics, which relate the probability of black hole formation to the initial conditions and the non-linear dynamics involved. Within a given statistical framework, further ambiguities arise—such as the eventual choice of a window function, the threshold criterion that an estimator must exceed to trigger collapse, and the numerical coefficients used in the computation (inferred, for instance, from numerical simulations \cite{Musco:2004ak, Escriva:2019nsa,Musco:2025zft}). These elements, whether explicitly chosen or inherently defined by the framework, collectively have a substantial impact on the resulting PBH abundance, e.g.  \cite{Germani:2018jgr, Musco:2018rwt, Ando_2018,Yoo_2018,Young:2024jsu,Fumagalli:2024kgg,Pi:2024ert}. Furthermore, PBHs are challenging to search for, meaning that theoretical predictions of their mass distribution are valuable in order to better inform and refine observational approaches.

Motivated by this, we present \path{pbhstat} as a simple tool that allows consistent computation of PBH abundance across different methods for ease of comparison. We aim for this tool to be accessible to non-experts who may not work directly on PBH formation, by providing an easy way to generate PBH mass functions from pre-defined primordial power spectra using consistent statistical formalisms. At the same time, it is designed to support experts in the field, allowing them to input custom power spectra and make detailed choices within various statistical approaches. While many researchers have developed private implementations of these methods, to the best of our knowledge, they have not yet been integrated into a publicly available, unified codebase.
We hope that this package will remove the need to re-implement standard techniques and serve as a platform for reproducibility and cross-checks across different approaches.

We implement three choices of statistics to describe how fluctuations collapse to form PBHs. Press-Schechter theory \cite{press_original,Carr:1975qj} provides a relatively simple method based on the probability that the density contrast exceeds a critical threshold. It usually assumes a Gaussian distribution for the linear overdensity and integrates over all regions above this threshold to estimate the collapse fraction, offering a first-order approximation that is easy to implement. Peaks theory \cite{peaks_original} {was first applied to the PBH context in the mid-2000s in \cite{Green:2004wb}}, {going beyond Press-Schechter} by modelling PBHs as forming from rare, high peaks in the curvature perturbation field. {It is generally considered to be a more accurate description of PBH formation and has been explored and developed extensively in the literature, see e.g.~\cite{Young_2014,Germani:2018jgr,Young:2020xmk,Young:2024jsu}.} Finally, we place particular emphasis on the more involved implementation of the non-linear compaction function statistics \cite{Fumagalli:2024kgg,Germani:2019zez,Germani:2023ojx}. This approach relies on using the initial conditions of the compaction function as the primary statistical measure. It focuses on identifying regions where the overdensity exhibits a pronounced peak, the compaction function attains its maximum at a characteristic smoothing scale relative to this peak, and this maximum surpasses the critical threshold necessary for the initial overdensity to collapse to form a black hole.
We do not delve into the underlying physics of the statistical methodologies used and instead refer the reader to the relevant references for detailed discussions.

Under the assumption of nearly Gaussian initial conditions, all methods depend exclusively on the initial power spectrum. The latter should be motivated by a realisable inflationary scenario, see e.g. \cite{_zsoy_2023} for a review on inflationary mechanisms that lead to PBH production.

Additional effects—such as quantum diffusion \cite{Pattison_2017,2018JCAP...08..018E,Biagetti:2021eep,Vennin_2025} and non-Gaussianities \cite{Bullock:1996at, Ivanov:1997ia, Young:2013oia, Byrnes:2012yx, Tada:2015noa, Atal:2018neu, Franciolini:2018vbk,Kehagias:2019eil,Riccardi:2021rlf,Gow:2022jfb,Matsubara_2022}—may play a crucial role in shaping the final abundance. Nonetheless, even when one adopts the simplifying assumption that the primordial curvature perturbations are Gaussian with a known power spectrum, the present tool provides useful first-order calculations of the mass function, illustrating the number of components that must be taken into account in different statistical approaches. We envisage implementing additional refinements in future versions of the code.

This companion to the code is organised as follows: in \cref{sec:output} we outline the main output of the code, in \cref{sec:stats} we present the three statistical methods for calculating PBH abundance that we implement, along with the primordial power spectra and window functions in \cref{sec:pps}. We lay out the modular structure of the code in \cref{sec:structure} and provide installation and quick-start usage instructions in \cref{sec:usage}. We comment on future additions and summarise in \cref{sec:conc}.

\section{Output of the Code}\label{sec:output}
The main output of \path{pbhstat} is the mass function of primordial black holes $f(M)$, which represents the fraction of energy density contained in primordial black holes per logarithmic mass interval:
\begin{equation}\label{eq:fm}
    f(M_{})=\frac{1}{\Omega_{\rm CDM}}\frac{d\Omega_{\rm PBH}}{d\ln M}
\end{equation}
which satisfies
\begin{equation}
    \int f(M) d\ln M= f_{\rm PBH}\equiv\Omega_{\rm PBH}/\Omega_{\rm CDM}, 
\end{equation}
where {$M$ is the mass of the PBH}, $f_{\rm PBH}$ is the fraction of dark matter energy density contained in PBHs today and $\Omega_{\rm CDM}$ is the energy density in cold dark matter today. Various definitions of the mass function are present in the literature, often simply related to the above one by a rescaling or numerical coefficients.

The PBH abundance today, $\Omega_{\rm PBH}$, can be obtained by integrating the contributions over all scales:  
\begin{equation}\label{eq:OmPBH}
    \Omega_{\rm PBH} = \int d\ln R\, \left( \frac{R_{\rm eq}}{R} \right) \beta(R),
\end{equation}
where $R$ is the physical scale of the horizon at the time of formation, which is related to the horizon mass, $M_{\rm H}$, as follows \cite{Nakama_2017}:
\begin{equation}\label{eq:RMh}
    R = \left(\frac{M_H}{M_{\rm eq}}\right)^{1/2} R_{\rm eq}.
\end{equation}
$\beta(R)$ is the mass fraction of PBHs at the time of formation, i.e. $\beta=\rho_{\rm PBH}/\rho_{\rm tot}|_{\rm formation}$. The factor $(\Meq/ M_H)^{1/2}$ scales $\beta$ to its value at matter-radiation equality exploiting the fact that during radiation domination $\beta\propto \rho_{PBH}/\rho_R \propto a $, and $\aeq/a(M_H) = (\Meq/M_H)^{1/2}$.
We use $M_{\mathrm eq}=2.8 \times 10^{17} M_\odot$ and {$k_{\mathrm eq}=1/R_{\mathrm eq}=0.01(\Omega_m/0.31)\,{\rm Mpc^{-1}}$, neglecting} for simplicity the small correction due to the evolution of the number of relativistic degrees of freedom in \cref{eq:RMh}. In practice, the integral over all scales in \cref{eq:OmPBH} will be computed over horizon mass, where it is best to integrate between extrema that correspond to the scales where the primordial power spectrum has decreased by a factor of 10 from its peak amplitude.
The implementation of $\beta$ has a different form depending on how the perturbations are assumed to collapse. We provide three methods from the literature for calculating $\beta$ that assume three different statistics, namely Press-Schechter, peaks theory and non-linear compaction statistics. The key ingredient to $\beta(R)$ is the primordial power spectrum, which can be either analytically calculated or loaded as a bespoke spectrum in the package. A separate module smooths the power spectrum and calculates the relevant width parameters $\sigma(R)$ which are fed to $\beta(R)$ as described in \cref{sec:pps}. To finally arrive at the PBH mass function, one has to invert the expression for the PBH mass in terms of the horizon mass and the statistical estimator chosen (see below).

\section{Choice of Statistics}\label{sec:stats}

In all three statistics methods, we use the compaction function (in its linear or non-linear version in $\zeta$) as an estimator.
The compaction function, denoted with $C$, measures the mass excess within a given areal radius \cite{Shibata:1999zs}. In the long-wavelength approximation, and by exploiting the (approximate) spherical symmetry of high peaks in a randomly distributed Gaussian field (in this case, $\zeta$) \cite{peaks_original}, the time-independent super-horizon compaction function can be written as \cite{Harada:2015yda}:
\begin{equation}\label{eq:compaction}
C(R) = g(R)\left(1-\frac{3}{8}g(R)\right),
\end{equation}
where
\begin{equation}\label{definitions1}
g = -\frac{4}{3}R\partial_R\zeta(R)
\end{equation}
which is sometimes referred to as the linear compaction function.\footnote{In some works, this variable is also denoted by $C_G$, $C_\zeta$, or $\delta_l$.}
Here, $\zeta(R)$ denotes the radial profile of the comoving curvature perturbation in a coordinate system centered on one of its peaks.
The compaction function $C$ on super-horizon scales is bounded from above, with $C \leq 2/3$. This bound is reached for $g = 4/3$, which separates standard Type~I fluctuations ($g \leq 4/3$) from more exotic Type~II fluctuations ($g > 4/3$), the latter being characterised by a non-monotonic areal radius. The current implementation of the code considers only Type~I fluctuations. Critical collapse for Type~II fluctuations is currently under investigation--see \cite{Uehara:2024yyp,Shimada:2024eec,Inui:2024fgk,Escriva:2025rja} and a dedicated statistical framework should be developed to properly explore this case. Accordingly, we bound our integral in $g$ to $4/3$. In practice, this consideration is relevant primarily for the non-linear statistics discussed in \cref{sec:nonlinear}. For the statistics in \cref{sec:PS} and \cref{sec:BBKS}, the constant threshold is sufficiently far from $4/3$ that Type~II fluctuations are naturally excluded.
\subsection{Press-Schechter}\label{sec:PS}
We implement Press-Schechter statistics in the \path{PressSchechterModel} class within the \path{collapse_stats.py} module following \cite{Gow:2020bzo} (and references therein). Note that in the original Press–Schechter formalism for PBHs, the mass fraction is typically expressed as an integral over the density contrast $\delta$, which is assumed to be Gaussian. When the time-independent component of the linear density contrast is averaged over a volume enclosed by a sphere of radius $R$, it becomes equivalent to the compaction function estimator $C$ (see for instance \cite{Young:2019osy}) on super-horizon scales. This is valid for the linear (in $\zeta$) term of the compaction function, i.e. $g$, and the linear smoothed density contrast. 
Using $g$ and adjusting the critical threshold $C_c$ accordingly \cite{Gow:2020bzo} means that we can write the subsequent expressions in terms of $g$ in place of the original $\delta$.

The form of $\beta$ in terms of the horizon mass (which can be converted from $R$ using \cref{eq:RMh}) is
\begin{equation}\label{eq:betapress}
    \beta(M_H)=2\int^{4/3}_{\C_c}d\C \frac{M}{M_H}P(\C)
\end{equation}
with $P(\C)$ the probability distribution function of the initial density perturbations and where the upper limit of the integral comes from excluding Type II fluctuations. Under the Gaussian approximation this reads
\begin{equation}\label{eq:pdf}
    P(\C) = \frac{1}{\sqrt{2\pi \sigma^2(M_H})}\exp\left(-\frac{\C^2}{2\sigma^2(M_H)}\right)
\end{equation}
where $\sigma(M_H)$ is the mass variance, which corresponds to the zero-th width parameter $\sigma_0$ in the hierarchy of moments\footnote{Here we defined the width parameters to be dimensionless.}
\begin{equation}\label{eq:sigs}
    \sigma_n^2 =\frac{16}{81}\int^\infty_0 \frac{dk}{k} (kR)^{4+2n} W^2(k,R) \mathcal{P}_\zeta\end{equation}
where $(kR)^4$ accounts for the super-horizon growth of the density perturbations, described by the dimensionless primordial power spectrum of the comoving curvature perturbation $\mathcal{P}_\zeta$, defined by
$\langle\zeta_\mathbf{k}\zeta_{\mathbf{k}'}\rangle = (2\pi)^3\delta(\mathbf{k}+\mathbf{k}')(2\pi^2/k^3) {\cal P}_\zeta(k)
$, and $W(k,R)$ is a window function chosen
to filter the perturbations over the smoothing scale $R$.

The window functions implemented for use with Press-Schechter are the Gaussian window function\footnote{This choice, strictly speaking, refers to a non-canonical definition of the compaction function; see \cite{Young:2020xmk}.
} and the real-space top-hat window function. 
A real-space top-hat window corresponds to averaging over a sharp spherical region of radius $R$
\begin{equation}
W_{\rm TH}(kR) = 3 \frac{ \sin(kR) - kR \cos(kR)}{(kR)^3}
\end{equation}
while the Gaussian window function corresponds to a weighted average over all space, where points closer to the center are weighted more heavily, and the weight decays exponentially with distance
\begin{equation}
W_{\rm G}(kR) = e^{-x^2/4}
\end{equation}{where we have defined $x\equiv kR$.} 
{From \cref{definitions1}, one can show \cite{Germani:2019zez,Germani:2023ojx} that $\sigma^2_g\equiv\langle g g \rangle = \sigma_0^2$ in \cref{eq:sigs}, with the window function automatically given by the real-space top-hat one. }

The critical scaling behaviour of the PBH mass can be written as \cite{PhysRevLett.80.5481}
\begin{equation}\label{critical1}
M=K M_H(\C-\C_c)^{\gammacr},
\end{equation}
where $K$, $\gammacr$ and $\C_c$ are numerical coefficients that should be chosen to be consistent with the smoothing function used to calculate $\sigma(M_H)$. Standard choices based on fiducial profiles are $\{K = 4,\; \C_c = 0.77\}$ for the real-space top-hat window function \cite{Young:2019osy} and $\{K = 10,\; \C_c = 0.28\}$ for the Gaussian window function \cite{Young:2020xmk}. These values of $\C_c$ are obtained by solving \cref{eq:compaction} for $\C$ with $C = 0.55$ for the top-hat, and $C = 0.25$ for the Gaussian window functions respectively, so as to account for the fact that we are using only the linear term of the compaction function \cite{Young:2019yug}.
 $\gammacr$ is implemented as an optional argument with default value 0.36 in the \path{MassVariance} class within the \path{mass_variance.py} module.
As is customary in these statistics, we do not consider here the profile dependence of the threshold.
We note that for the real-space top-hat window function, the integral over $\beta(R)$ diverges and therefore we implement a large-$k$ cut-off as an optional argument in the \path{MassVariance} class, following \cite{Gow:2020bzo}. This is placed at the point where the window function reaches its first trough, at $k=4.49/R$. This removes the divergence from the integral, but the window function is no longer a true top-hat function.

Combining \cref{eq:betapress} and \cref{eq:pdf}, and trading the variable $g$ for $M$ by inverting \cref{critical1}, we can write $f(M)$ as:
\begin{equation}
f(M) =  \frac{1}{\Omega_{\rm CDM}} \int \mathrm{d}M_H \;
\sqrt{\frac{M_{\rm eq}}{M_H}} 
\frac{\left(\frac{M}{KM_H}\right)^{1/\gammacr}}{\gammacr M_H^2} 
M
P\left(\C_\bullet\right).
\end{equation}
where $\C_\bullet$ denotes $\C$ expressed in terms of the other variables by inverting the critical scaling relation in \cref{critical1}:
\begin{equation}\label{gbullet}
g_\bullet \equiv g(M,M_H) =\left(\frac{M}{K M_H}\right)^{1/\gammacr} +g_c.
\end{equation}

\subsection{Peaks Theory}\label{sec:BBKS}
We implement peaks theory statistics \cite{peaks_original,Green:2004wb} in the \path{PeaksTheoryModel} class within the \path{collapse_stats.py} module in the following way. The mass fraction is given by 
\begin{equation}\label{eq:betapeaks}
    \beta(M_H) = b\int^{4/3}_{\C_c}d\C\,\frac{M}{M_H}n\left(\frac{\C}{\sigma(M_H)}\right)
\end{equation}
where $b$ is a pre-factor that accounts for the volume over which the window function is integrated, taking $b=\{(2\pi)^\frac{3}{2}, 4\pi/3\}$ for Gaussian and top-hat window functions respectively \cite{Young:2024jsu}. $n$ is the number density of peaks given by
\begin{equation}\label{eq:npeaks}
    n\left(\C,M_H\right) = \frac{1}{3^\frac{3}{2}(2\pi)^2}\left(\frac{\sigma_1(M_H)}{\sigma_0(M_H)}\right)^3\left(\frac{\C}{\sigma_0(M_H)}\right)^3\exp\left(-\frac{1}{2}\left(\frac{\C}{\sigma_0(M_H)}\right)^2\right).
\end{equation}
with $\sigma_0$ and $\sigma_1$ as defined in \cref{eq:sigs}. Note that we have absorbed the usual $R^3$ factor in this equation into the dimensionless expression for $\sigma_1$, see \cref{eq:sigs}.

The window functions implemented for use with the \path{PeakTheoryModel} are the same as for the \texttt{PressSchechterModel}, namely the Gaussian window function and the real-space top-hat window function. The recommended choices for the numerical coefficients are the same as in the previous \cref{sec:PS}. The large-$k$ cut-off flag is also compatible with this statistics choice and can be implemented in the \path{MassVariance} class within the \path{mass_variance.py} module.

Putting \cref{eq:betapeaks} and \cref{eq:npeaks} together with \cref{eq:fm}, we compute the mass function with \cite{Gow:2020bzo}
\begin{equation}
f(M) = \frac{b}{2}\frac{1}{\Omega_{\rm CDM}} \frac{1}{\gammacr K^\frac{1}{\gammacr}}\int dM_H \sqrt\frac{M_{\rm eq}}{M_{\rm H}^3}\left(\frac{M} {M_H}\right)^{1 + \frac{1}{\gammacr}} n\left(\C_\bullet, M_H\right),
\end{equation}
where again $g_\bullet$ is given in \cref{gbullet} by inverting the critical scaling relation in \cref{critical1}.

\subsection{Non-Linear Compaction Function Statistics}\label{sec:nonlinear}

We implement non-linear compaction function statistics \cite{Germani:2019zez} in the \path{NonLinearModel} class within the \path{collapse_stats.py} module. % in the following way.
For this method, the estimator for the PBH abundance is the full super-horizon compaction function $C$ in \cref{eq:compaction}. Since PBH formation occurs when the super-horizon $C$ at its maximum ($R = R_m$) exceeds a certain threshold $C_c$ \cite{Musco:2018rwt}, it is useful to introduce the following variables:
\begin{equation}\label{definition2}
    \qquad v = R g',\qquad w = -R^2 g'',
\end{equation}
where primes denote derivatives with respect to the comoving radial coordinate $R$. In the following expressions, these quantities are understood to be evaluated at the maximum of $C$.

By requiring that these three conditions are satisfied — (i) the overdensity has a peak, (ii) the compaction function reaches a maximum ($v=0$ and $w \geq 0$), and (iii) this maximum exceeds a critical threshold ($g\geq g_c$) — one can rewrite $\beta(R)$ in the following convenient form \cite{Germani:2023ojx} (see also \cite{Fumagalli:2024kgg}):
\begin{equation} \label{beta2}
\beta(M_H) = \frac{2\pi}{3}\int_{0}^{\infty} dw w \int_{g_c(w)}^{4/3} dg\,  \frac{f\left(\frac{\chi}{\sigma_\chi}\right)}{(2\pi )^{3/2}(\sqrt{3}\sigma_1/\sigma_2)} \frac{M}{M_H} P(g,w,v=0),
\end{equation}
where $\chi = 2g + w$ is the trace of the Hessian at the position of the peak's center under the condition $v=0$.
It should be noted that, here and in the following, $(g,w)$ are evaluated at the maximum of the compaction function. The explicit form of the function $f$, arising from condition (i), is given in Equation (A15) of Ref. \cite{peaks_original}. 
The probability distribution function $P(g,w,v=0)$, obtained from the three conditions mentioned above, is given by
\begin{equation}
\label{Pgwv}
P(g,w,v=0) =p(v=0)p(g,w)=  \frac{1}{\sqrt{2\pi \sigma_v^2}}p(g,w), 
\end{equation}
where $p(v=0)$ is the (assumed) Gaussian distribution of $v$ evaluated at $v=0$, i.e. the probability of having a peak in the compaction function, multiplied by the conditional joint probability of the other two variables $p(g,w)$. Assuming again for simplicity that primordial fluctuations, and therefore $g$, are Gaussian, $p(g,w)$ becomes a bi-variate Gaussian distribution for the two correlated variable $w$ and $g$ with an extra correction inherited from imposing the condition $v=0$. We then have (using $\sigma_{x}^2\equiv\langle x^2\rangle$, $\sigma_{xy}\equiv \langle xy\rangle$, 
$\gamma_{xy}\equiv\frac{\sigma_{xy}}{\sigma_{x}\sigma_{y}}$):
\begin{equation}
p(w,g) = \frac{1}{2\pi \sqrt{\det \Sigma}} \exp\left(-\frac{1}{2} \vec{X}^T\Sigma^{-1} \vec{X}\right),
\end{equation}
where 
\begin{equation}
\vec{X}^T = (w\,\,g), \quad \Sigma = 
\begin{pmatrix}
\tilde{\sigma}^2_{w} & \tilde{\sigma}^2_{w g}\\
\tilde{\sigma}^2_{g w} & \tilde{\sigma}^2_{g},\\
\end{pmatrix},
\end{equation}
with
\begin{align}\label{sigmatildes}
\tilde{\sigma}^2_g = \sigma^2_g(1 - \gamma_{v g}^2),\quad \tilde{\sigma}^2_w = \sigma^2_w(1 - \gamma_{v w}^2),\quad
\tilde{\sigma}^2_{w g}  = \sigma^2_{w g} - \frac{\sigma_{w v}^2 \sigma_{v g}^2}{\sigma^2_v }.
\end{align}
The probability distribution function can therefore be written explicitly as
\begin{align}\label{pdfs}
p(g,w)&= \frac{1}{\sqrt{2\pi\tilde{\sigma}_w^2 (1-\tilde{\gamma}^2)}} \exp\left(-\frac{(w -  \tilde{\gamma}\frac{\tilde{\sigma}_w}{\tilde{\sigma}_g} g)^2 }{2\tilde{\sigma}_w^2(1-\tilde{\gamma}^2)}  \right)  \frac{1}{\sqrt{2\pi \tilde{\sigma}_g^2}}\exp\left(-\frac{g^2}{2\tilde{\sigma}^2_g}\right),
\end{align}
or analogously exchanging $g$ with $w$. $\tilde{\gamma}$, the so-called Pearsol coefficient, tells us the strength of the cross-correlation between the two variables:
\begin{equation}
\tilde{\gamma} = \frac{\tilde{\sigma}_{w g}^2}{\tilde{\sigma}_w \tilde{\sigma}_g}.
\end{equation}
The threshold depends on the shape of the compaction function and is characterised by the function $g_c(w)$, where the dimensionless curvature of the linear compaction function $w$ parametrises the profile. 
In the current implementation, we adopt the threshold derived using the ``q-approach'' \cite{Escriva:2019phb}, which asymptotically approaches $4/3$ for large $w$, corresponding to sharply peaked compaction function profiles. This method relies on fiducial $C$ profiles expressed in terms of the normalised variable $q \equiv -R^2 C''/[4C\!(1 - \tfrac{3}{2}C)].$
One first inverts the universal threshold on the average compaction function \cite{Escriva:2019phb,Kehagias:2024kgk} to obtain the threshold of the compaction function as a function of $q$:
\begin{equation}\label{Ccq}
C_c(q)= \frac{4}{15}e^{-1/q}\frac{q^{1-\tfrac{5}{2q}}}{\Gamma\left(  \tfrac{5}{2q}\right) -\Gamma\left(  \tfrac{5}{2q},\tfrac{1}{q}\right)},
\end{equation}
where $\Gamma(x)$ and $\Gamma(x,y)$ denote the Euler gamma function and the upper incomplete gamma function, respectively. Then, to express the threshold as a function of $w$, one uses the relation\footnote{We remind the reader} that all expressions are considered at the maximum of the compaction function  where $C' = g' = 0$.
\begin{equation}\label{wq}
w = -R^2 C'' = 4 q\, C\, \sqrt{1 - \frac{3}{2} C}.
\end{equation}
Evaluating the right-hand side of the previous expression at the threshold $C_c$ and substituting \eqref{Ccq}, one finds $q(w)$. Finally, this expression is substituted into the inverted relation for $g$ as a function of $C$ to obtain the threshold for $g$ as a function of $w$:
\begin{equation}\label{gcw}
g_c(w) = \frac{4}{3} \Bigg( 1 - \sqrt{1 - \frac{3}{2} C_c(q(w))} \Bigg).
\end{equation}
The steps outlined in Eqs. \eqref{Ccq}, \eqref{wq} and \eqref{gcw} are implemented in the module \path{stats_utils.py} to compute $g_c(w)$.

Within this method it is customary to express the mass of the PBH using the critical scaling given in terms of the compaction function as
\begin{equation}\label{criticalscaling}
M=\tilde{K} M_H\left[C(g) - C(g_c(w)) \right]^{\gammacr}.
\end{equation}
The term in parentheses, when expanded around $g_c$, reduces to \cref{critical1} with a different re-scaled coefficient which we denote by $\tilde{K}$.
The two numerical coefficients are optional arguments in the code, but we choose $K= 6$, and critical exponent $ \gammacr= 0.36$ as default values \cite{Escriva:2019phb}.

The final abundance depends crucially on the choice of the profile dependent threshold $ g_c(w) $ and to some extent, the critical scaling. Both elements are subject to ongoing scrutiny in the literature \cite{Uehara:2024yyp,Shimada:2024eec,Inui:2024fgk,Escriva:2025rja} and can be readily modified within the \path{stats_utils.py} module of the code in the future.

Inverting \eqref{criticalscaling} and trading the variable $g$ for $M$, one can derive the mass function \cite{Fumagalli:2024kgg}:
\begin{equation}\label{fPBH}
f(M) = \frac{2\pi \tilde{K}}{3}\int d\ln M_H\int_0^{\infty} dw \frac{w \,(\frac{M}{K M_H})^{\frac{1}{\gammacr} +1}}{\gammacr (1-\frac{3}{4}g_\bullet)}\left(\frac{\Meq}{M_H}\right)^{1/2} P(g_\bullet,w,v=0)\frac{f\left(\frac{2g_\bullet+w}{\sigma_\chi}\right)}{(2\pi )^{3/2}(\sqrt{3}\sigma_1/\sigma_2)}.
\end{equation}
Here, $g_\bullet$ denotes $g$ evaluated by inverting \cref{criticalscaling}.

For completeness, we report the expressions for the various correlators used in this choice of statistics. All of them can be derived from the definitions of the relevant variables in \cref{definitions1,definition2}, the $\sigma$’s defined in \cref{eq:sigs}, and the standard expression for the dimensionless primordial power spectrum given just below that equation:
\begin{equation}\label{sigmas2}
\begin{aligned}
    \sigma_g^2 &=\sigma_0^2, \qquad \sigma_w^2 = \sigma_2^2 -4\sigma_1^2 +4\sigma_0^2,\qquad  \sigma_\chi^2 =\sigma_2^2\\[1mm]
    \sigma^2_{v} &= \frac{16}{81}\int \frac{dk}{k} \left(3x\sin x-x^2 W(x)\right)^2 \,\mathcal{P}_{\zeta}\\[1mm]
\sigma_{vg}&\equiv \langle v g\rangle =\frac{16}{81} \int^\infty_0 \, \frac{dk}{k} \left(3 \, x \sin x - x^2 \, W(x)\right) \, W(x) \, x^2 \, \mathcal{P}_{\zeta} \\[1mm]
\sigma_{gw} &= \frac{16}{81} \int^\infty_0 \, \frac{dk}{k} \left(x^6 - 2 \, x^4\right) \, W^2(x) \, \mathcal{P}_\zeta \\[1mm]
\sigma_{vw} &= \frac{16}{81} \int^\infty_0 \, \frac{dk}{k} \left(3 \, x \sin x - x^2 \, W(x)\right) \left(\, x^4 - 2 \,  \, x^2\right)W(x) \mathcal{P}_\zeta,
\end{aligned}
\end{equation}
where we have defined $x\equiv kR$.
Note that, if one strictly takes as an estimator the orthodox definition of the compaction function $C$ (as in \cref{eq:compaction}), then $W$, playing the role of the window function in the formula above, is automatically the real-space top-hat function. 

In practice, we choose the range of $k$-values so that only those corresponding to amplitudes of the primordial power spectrum within a factor of 10 of the peak amplitude are integrated over. In this way, the code determines an effective width of the power spectrum and sets the limits of both $w$ (see~\cite{Fumagalli:2024kgg}) and $M_H \sim R^2 \sim k^{-2}$ accordingly. This \path{threshold_factor} can be modified from its default value of 0.1 in the \path{collapse_stats.py} module. We emphasise that it may be particularly important to test different values if the input power spectrum is especially broad, to ensure convergence of the $\sigma$’s in \cref{sigmas2} and to properly account for high values of $w$.

As already mentioned, if initial non-Gaussianity is assumed to be present, the expressions in this \cref{sec:stats} have to be corrected with the specific form of non-Gaussianity included, see e.g. \cite{Riccardi:2021rlf, Gow:2022jfb,Matsubara_2022,Kehagias:2019eil}. 

\section{Power Spectra and Smoothing}\label{sec:pps}

We implement both pre-set options for the primordial power spectrum, as well as the possibility for a user-defined primordial power spectrum that can be defined analytically or by uploading an array of $\mathcal{P}_\mathcal{R}$ along with their corresponding wavenumbers $k$.

The key preset options are \path{piecewise}, defined as
\begin{equation}
    \mathcal{P}_\mathcal{R}(k) = 
\begin{cases}
A\left( \dfrac{k}{k_\star} \right)^{n_g}, & \text{if } k \leq k_\star \\
A\left( \dfrac{k}{k_\star} \right)^{-n_d}, & \text{if } k > k_\star
\end{cases}
\end{equation}
where $A$ is the amplitude of the peak of the power spectrum, $k_\star$ is the position of the peak in the power spectrum, $n_g$ is the index of the growing slope and $n_d\geq0$ is the index of the decaying slope. We also set the minimum amplitude of the power spectrum at any $k$ to be $A=2\times10^{-9}$ in order to match CMB observations at large scales. While such small amplitude perturbations will not affect PBH abundance, this setting can be overridden by adjusting the value of \path{min_amplitude} in the \path{power_spectrum.py} module. 

The \path{lognormal} power spectrum class is defined by
\begin{equation}
\mathcal{P}_\mathcal{R}(k) = A \exp\left( -\frac{1}{2} \left( \frac{\ln(k / k_\star)}{\sigma_{\rm ln}} \right)^2 \right)
\end{equation}
with $A$ the amplitude of the peak, $k_\star$ the position of the peak and $\sigma_{\rm ln}$ the width of the log-normal distribution. We also implement a \path{delta} class by setting $\sigma_{\rm ln}$ to be very small (e.g. 0.001). This produces a very narrow log-normal distribution which can be used to approximate a delta-function and avoids numerical issues with using a true delta-function in the same framework as the rest of the code. We note that this type of power spectrum is unphysical but it is often used for simplicity.

The \path{flat} power spectrum class is scale-invariant across the range of scales provided by the user, simply defined with an amplitude $A$.

These pre-set power spectra aim to capture the key classes of inflationary models in the literature that lead to PBH production. We emphasise that a numerical power spectrum calculated by evolving the equations of motion for the inflaton with a given inflationary potential should be used in order to calculate the precise PBH abundance from specific initial conditions.

Finally, the \path{custom} class allows the user to load an array of $\mathcal{P}_\mathcal{R}$ values that correspond to an array of $k$-values.

The values of $\mathcal{P}_\mathcal{R}$ are fed to the \path{MassVariance} class. The calculation of the relevant $\sigma(R)$, depending on the choice of statistics, involves convolving $\mathcal{P}_\mathcal{R}$ with the chosen window function to find the smoothed power spectrum for Press-Schechter and peaks theory methods, as well as other combinations for the non-linear statistics. These expressions are then integrated over the user-defined $k$-range such that the output of the \path{MassVariance} class is an interpolating function for the relevant $\sigma(R)$, where the $R$-range of validity for the interpolators are given by $R=1/k$.

\section{Structure of the Code and Installation}\label{sec:structure}

The \path{pbhstat} package is organised into modular components, each handling a distinct stage of the calculation pipeline outlined above, from reading the initial primordial power spectrum to computing the PBH mass function. The core codebase resides in the \path{src/} directory, structured as follows:

\begin{itemize}
    \item \textbf{\path{power_spectrum.py}} \\
    Contains the \path{PowerSpectrum} class, which defines and evaluates the primordial curvature power spectrum. There are some pre-defined forms for the power spectrum which can be loaded, such as \path{flat}, and \path{piecewise}. It also supports both analytical and tabulated user-defined spectra which can be implemented directly into the class or loaded as an array. The class then provides interpolation over a user-specified range of wavenumbers $k$.

    \item \textbf{\path{mass_variance.py}} \\
    Implements the \path{MassVariance} class, which computes the smoothed mass variance $\sigma^2(R)$ over a range of scales, using the power spectrum and a choice of window function. Currently, the Gaussian window function and the real-space top hat window functions are implemented, with a large$-k$ cut-off optional for the latter.

    \item \textbf{\path{collapse_stats.py}} \\
    Provides statistical models for PBH formation. It includes classes for the Press--Schechter formalism (\path{PressSchechterModel}), peaks theory (\path{PeaksTheoryModel}), and for non-linear compaction statistics (\path{NonLinearModel}). These classes set up the computation of the mass function of primordial black holes as a function of mass.

    \item \textbf{\path{mass_function.py}} \\
    Contains the \path{MassFunction} class, which wraps around the collapse statistics to compute the differential PBH mass function $f(M)$ and total abundance $f_{\rm PBH}$.

    \item \textbf{\path{stats_utils.py}} \\
    Hosts auxiliary functions used predominantly for the \path{NonLinearkModel} class, such as critical threshold calculations, filtering functions, and intermediate formulae that can be pre-computed.

    \item \textbf{\path{plot_utils.py}} \\
    Provides helper routines for plotting key quantities such as the power spectrum, mass variance, and PBH mass function.

    \item \textbf{\path{constants.py}} \\
    Stores fiducial values of cosmological quantities that can be edited and will be imported for abundance calculations.
    
\end{itemize}

This modular architecture is designed to be user-friendly and easily adaptable with extensions and further refinements as the user sees fit.

\section{Usage Instructions}\label{sec:usage}
\subsection{Installation}
To install, it is recommended to create either a Python virtual environment or a conda environment, and then run: 
\begin{verbatim}
pip install pbhstat
\end{verbatim}

We provide two example \path{jupyter} Python notebooks for quick-start examples of how to use the code when installed using \path{pip}. Alternatively, download the code from \path{github} with 
\begin{verbatim}git clone https://github.com/pipcole/pbhstat.git\end{verbatim} and run 
\begin{verbatim}cd pbhstat
pip install -e .\end{verbatim} for an editable install. The required dependencies are \path{numpy}, \path{matplotlib}, \path{scipy}, and \path{tqdm}.

The code has been tested with Python version 3.9.15 on an Apple M3 Pro running macOS Sonoma 14.1 and with Python 3.12.3 on a Lenovo Yoga 7 2-in-1 14IML9 running Ubuntu 24.04 LTS as well as with Python version 3.11 in a Google Colab notebook.

\subsection{Example usage}
We provide two example Python notebooks with the code release for calculating the mass function and PBH abundance from a custom primordial power spectrum, and from a pre-defined piecewise power spectrum. We annotate below the step-by-step code snippets to do this.

First, import \path{numpy}, \path{pickle} or similar if loading own arrays from file and \path{pbhstat}, assuming it has been successfully installed with \path{pip}. Then import the relevant classes and functions from the package. Note that the imports will need to be edited if the package is not installed with \path{pip}.
\begin{minted}[bgcolor=codebg, fontsize=\footnotesize]{python}
import numpy as np
import pickle

import pbhstat

from pbhstat.power_spectrum import PowerSpectrum
from pbhstat.mass_variance import MassVariance
from pbhstat.collapse_stats import PressSchechterModel, PeaksTheoryModel, BroadPeakModel
from pbhstat.mass_function import MassFunction
from pbhstat.plot_utils import plot_power_spectrum, plot_mass_variance, plot_mass_function
\end{minted}

Next, define a power spectrum and corresponding $k$-values. All inverse length units should be in $\rm Mpc^{-1}$. This could be a custom power spectrum either defined analytically, for example:
\begin{minted}[bgcolor=codebg, fontsize=\footnotesize]{python}
# Set up k values in [1/Mpc]
k_values = np.logspace(3, 7, 3000)

# Custom power spectrum 
Ak = 0.008
Deltak = 1
kpeak = 1e5 # [1/Mpc]
P_k_custom = Ak * np.exp(-0.5 * (1/(np.sqrt(2 * np.pi) * Deltak)) * (np.log(k_values / kpeak) / Deltak)**2)
\end{minted}
or loaded from file:
\begin{minted}[bgcolor=codebg, fontsize=\footnotesize]{python}
with open("path_to_k_array.pickle", "rb") as f:
    k_values = pickle.load(f)
with open("path_to_P_k_array.pickle", "rb") as f:
    P_k_custom = pickle.load(f)
\end{minted}
which should then be instantiated with
\begin{minted}[bgcolor=codebg, fontsize=\footnotesize]{python}
ps_custom = PowerSpectrum(
    shape='custom',
    k_values=k_values,
    P_k_values=P_k_custom
)
\end{minted}
Alternatively, the power spectrum can be generated from one of the pre-set options, for example a piecewise power spectrum which can be instantiated directly:
\begin{minted}[bgcolor=codebg, fontsize=\footnotesize]{python}
# Set up k values in [1/Mpc]
k_values = np.logspace(3, 7, 3000)

# Piecewise power spectrum
ps_piecewise = PowerSpectrum(
    shape='piecewise',
    amplitude=0.008,
    k_star=1e5,
    ng=4,
    nd=2,
    k_values=k_values
)
\end{minted}
We will now assume that we are using \path{ps_piecewise} for the remainder of these instructions. The next step is to instantiate the mass variance (and higher moments) with a choice of window function (\path{realtophat} or \path{gaussian}) and statistics (\path{press}, \path{peaks} or \path{nonlinear}). When choosing the real top-hat window function, there is also an option to use the large $k$ cut-off as described in \cref{sec:PS} - the default is False. There are some incompatible choices which will throw an exception, for example that the non-linear statistics must be chosen together with the real top-hat window function.

\begin{minted}[bgcolor=codebg, fontsize=\footnotesize]{python}
# Instantiate mass variance with choice of window function and statistics
mv_piecewise = MassVariance(window='realtophat', power_spectrum=ps_piecewise, statistics='nonlinear', 
                            cutoff=False)
\end{minted}
Finally, all that is left is to instantiate the mass function class with the same statistics model as the mass variance, and evaluate on a grid of $M$ values, calculating $f_{\rm PBH}$ as well if desired. Numerical values of $K$, $g_c$ and $\gamma$ can be defined by the user here as optional arguments when instantiating the mass function class, but recommended values are defaulted to. 

Furthermore, correlations with $v$ can be enabled or disabled in the non-linear compaction function statistic using the flag \path{vcorr}. By toggling \path{vcorr}, the user can assess the impact of consistently accounting for the conditional probability of $v=0$, i.e., the probability of being at an extremum of the compaction function. In particular, setting \path{vcorr = False} implies that all $\tilde{\sigma}$ in \cref{sigmatildes} reduce to their untilde versions. It has been shown analytically in~\cite{Fumagalli:2024kgg} that for a broad spectrum, these correlations are subleading for $R \sim 1/k_{\rm IR}$, where $k_{\rm IR}$ is the infrared scale associated with the broad power spectrum.

The number of $M$ values that $f(M)$ is calculated for can be adjusted from the default 50 with the optional argument \path{mpbh_vals} in the mass function \path{evaluate} call.

\begin{figure}
    \centering
    \includegraphics[width=0.75\linewidth]{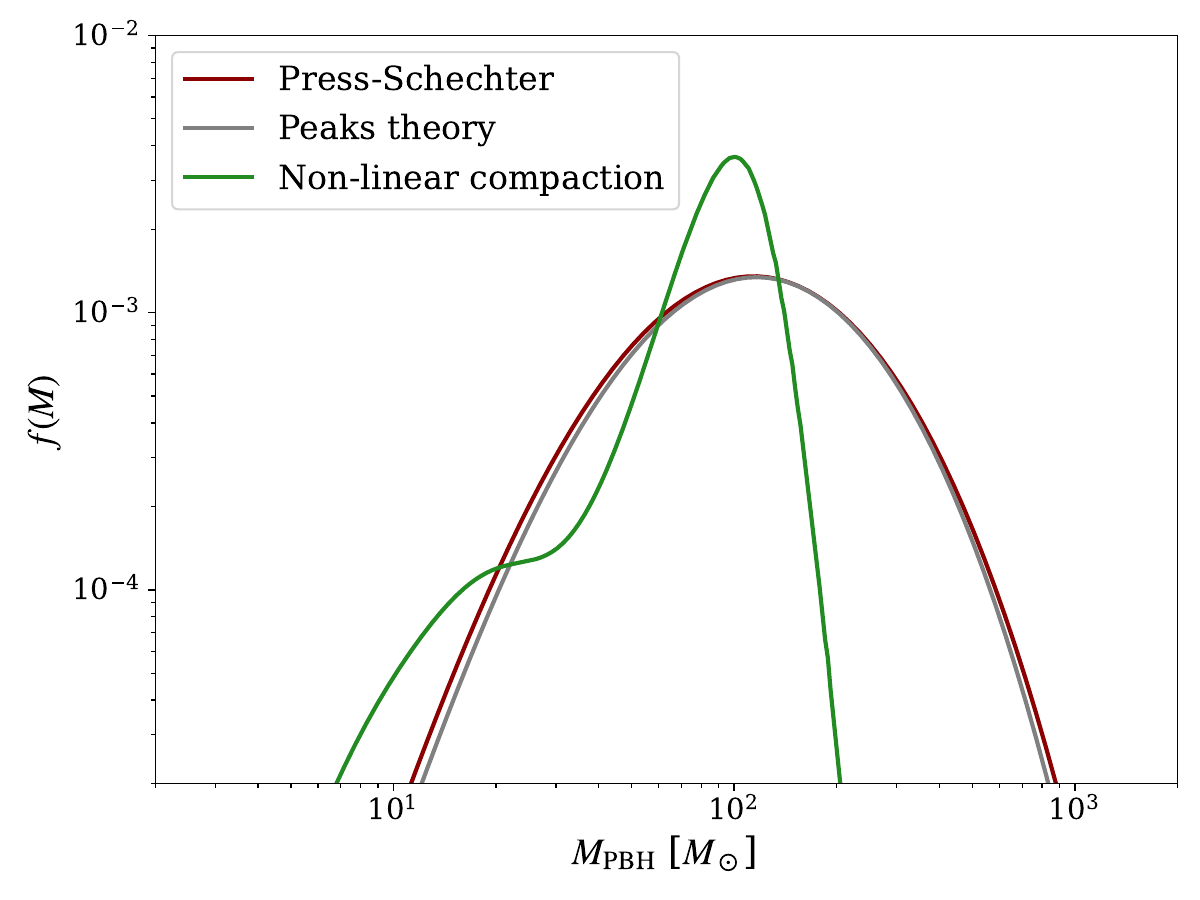}
    \caption{Comparison of mass functions calculated with the three different statistical methods. The power spectrum is chosen to be a log-normal distribution with width $\Delta=1$ and the window function is a real-space top hat. The amplitudes of the peak of the power spectrum are $A=\{ 0.00865, 0.0077, 0.009\}$ for Press-Schechter, peaks theory and the non-linear compaction statistics respectively. All mass functions result in $f_{\rm PBH}=2.5\times10^{-3}$.}
    \label{fig:fm_compare}
\end{figure}

\begin{figure}
    \centering
    \includegraphics[width=0.75\linewidth]{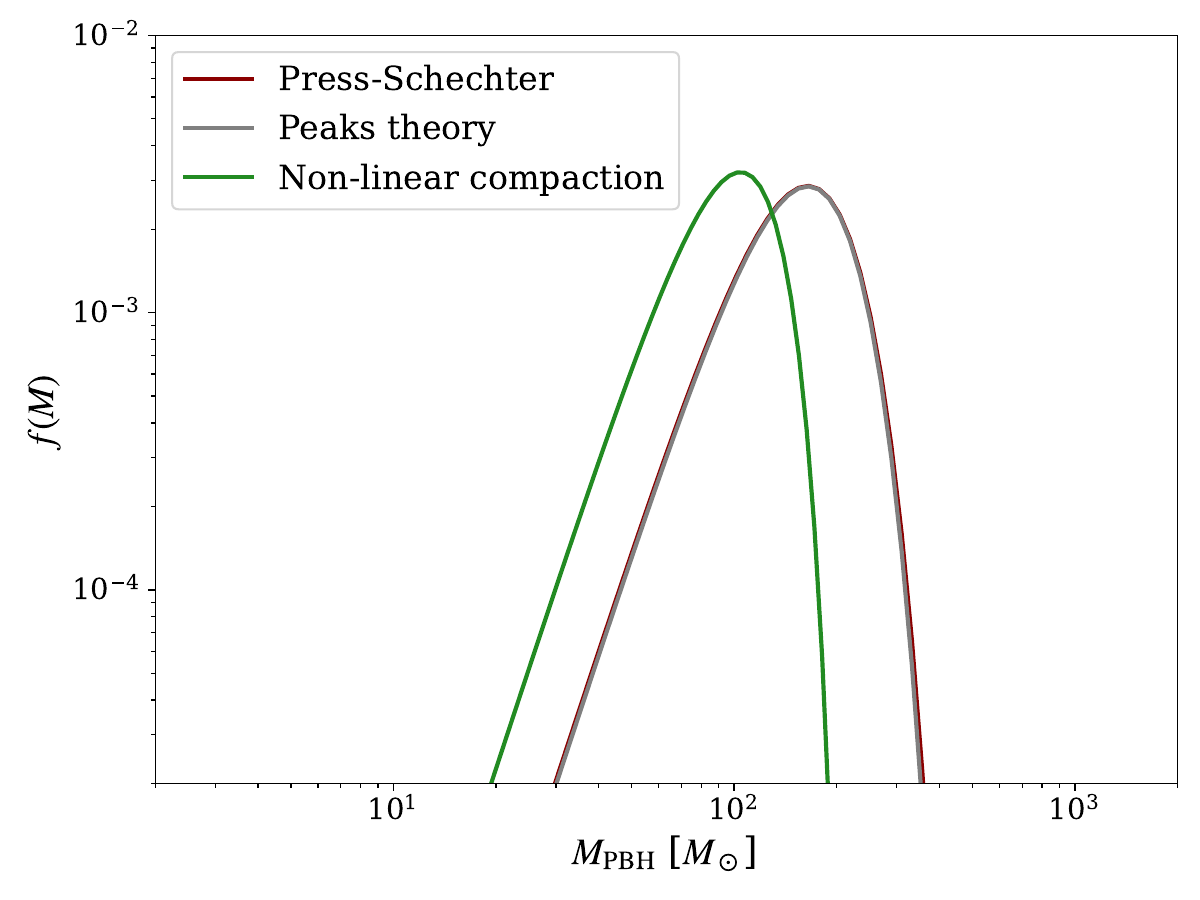}
    \caption{Comparison of mass functions calculated with the three different statistical methods. The power spectrum is chosen to be a narrow top-hat function with width in $k$-space equal to 1.3 and the window function is a real-space top hat. The amplitudes of the peak of the power spectrum are $A=\{0.02795, 0.02455, 0.03118\}$ for Press-Schechter, peaks theory and the non-linear compaction statistics respectively. All mass functions result in $f_{\rm PBH}=2.5\times10^{-3}$.}
    \label{fig:fm_comparenarrow}
\end{figure}

\begin{minted}[bgcolor=codebg, fontsize=\footnotesize]{python}
# Instantiate mass function
mass_function = MassFunction(mass_variance=mv_custom, statistics='nonlinear', K=6, vcorr=True, gamma=0.36)

# Evaluate f(M_PBH) and f_PBH
mpbh, f_mpbh = mass_function.evaluate(k_values, mpbh_vals=50)

fpbh = mass_function.fpbh(f_mpbh, mpbh)
\end{minted}
Key quantities can be plotted using the routines in \path{plot_utils.py}.
\begin{minted}[bgcolor=codebg, fontsize=\footnotesize]{python}
# Plot power spectrum
plot_power_spectrum(k_values, ps_piecewise(k_values))

# Plot mass variance
# Define R values
R_values = 1/k_values

# Evaluate mass variance at k_values
sigma0sq_piecewise = mv_piecewise.evaluate(k_values)

plot_mass_variance(R_values, sigma0sq_piecewise)

# Plot mass function
plot_mass_function(mpbh, f_mpbh, fpbh_val=fpbh)
\end{minted}
We plot a comparison of mass functions for different statistics choices in \cref{fig:fm_compare} for a broad log-normal distribution and in \cref{fig:fm_comparenarrow} for a narrow step function.

In \cref{fig:fm_compare}, the non-linear compaction function statistics yields a narrower and higher peak toward the infrared part of the spectrum compared to the other two statistical approaches considered. This behavior has been extensively discussed in Ref.~\cite{Fumagalli:2024kgg}, and we adopt the same underlying assumptions here, particularly regarding the threshold criterion. It was shown that for broad power spectra, the mass function is dominated by infrared modes associated with curvature profiles in which the compaction function is sharply peaked (i.e. with large values of $w$).
In the mildly broad log-normal example considered here, these infrared modes produce PBHs that are not-so-heavy due to the suppression implied by the critical scaling relation in \cref{criticalscaling}, namely $M \simeq M_H w^{-2\gammacr}$ for large $w$, resulting in PBHs with masses significantly below the horizon mass.
It is worth noting that while Press--Schechter and peaks theory lead to mass functions with the same shape, their overall amplitudes agree only because we choose the power spectrum amplitudes such that the total PBH abundance for the two cases are comparable. If instead the same value for the power spectrum amplitude is used for both methods, there is a difference of 2-3 orders of magnitude in the resulting PBH abundance as demonstrated in e.g. \cite{PhysRevD.104.083546}.
In \cref{fig:fm_comparenarrow}, as a benchmark for a narrow spectrum, we use a top-hat profile with width $k_{\rm UV}/k_{\rm IR} = 1.3$. For such narrow spectra, all statistics prefer the same comoving scale, namely $r \sim 2.74 / k_p$, which corresponds to the first maximum of $\sigma_g^2$. The difference in the position of the main peak in the resulting mass function arises from the different choices of threshold and overall normalisation in the critical scaling relation. Specifically, we adopt the same parameters for Press--Schechter and peaks theory (see \cref{sec:stats}), namely $(K, g_c) = (4, 0.77)$. For the non-linear compaction function statistics, we instead choose $\tilde{K} = 6$, and for a narrow spectrum, $g_c \simeq 0.91$ \cite{Fumagalli:2024kgg}.

In the case of a  monochromatic power spectrum, all methods produce mass functions that are nearly identical in shape because the statistical profile of the various estimators is essentially fixed. In contrast, for broader spectra, the situation becomes significantly more complex due to the influence of different statistical realisations. 
This influence is more relevant for the non-linear compaction function statistics, where one scans over different realisations of the curvature of the compaction function. Furthermore, the dependence on the threshold on $w$ provides a way to account for the influence of UV modes on IR modes. Whether this approach is sufficient is still under investigation. In contrast, for the other two statistics, this effect is incorporated by imposing an artificial cut-off at $k \sim 4.49 / R$ (see discussion in \cref{sec:PS}). These are the sources of the discrepancy behind the different curves in \cref{fig:fm_compare}. As stated in the introduction, the present tool introduced in this work is also intended to explore differences between distinct statistical approaches that remain under active investigation in the literature.

Finally, we provide a simple interface so that mass functions can be overlaid with the observational constraints available in the \path{PBHbounds} repository \cite{bradley_j_kavanagh_2019_3538999}, available here: \path{https://github.com/bradkav/PBHbounds/}.
Arrays of $f(M)$ and corresponding $M$ values can be saved to the directory in which \path{PBHbounds} is located with:
\begin{minted}[bgcolor=codebg, fontsize=\footnotesize]{python}
bounds_utility(mpbh, f_mpbh, 'path_to_PBHbounds_directory')
\end{minted}
Then, a modified version of the default script \path{PlotPBHbounds.py} is available at \path{https://github.com/pipcole/pbhstat} to be run in the directory of \path{PBHbounds}. This will overplot the mass function produced with \path{pbhstat} with the bounds (that can be specified via the command line, if not all bounds are desired) from \path{PBHbounds}. We emphasise that the constraints themselves depend on the form of the mass function, and therefore overlaying broad mass functions with monochromatic PBH constraints is inconsistent. We recommend that this utility should only be used for narrow mass functions, and in any case as an approximate guide. See \cite{Bellomo:2017zsr} for a method for converting constraints from monochromatic to extended mass functions.

\section{Summary and Outlook}\label{sec:conc}
\path{pbhstat} is a simple, user-friendly tool for computing the mass function of primordial black holes from an initial primordial power spectrum. Given the many sensitivities of this calculation to the choice of power spectrum, window function, numerical coefficients and statistics that describe the collapse of the perturbations, we aim to provide a community tool that enables easy comparisons.
%(see Ref.~\cite{xxx} within the LISA consortium).
%} {Here if you like we may add a reference: Author: LISA Cosmology working group, "PrimBHoles, a code to compute the GW signatures of primordial black holes", in preparation.} \pc{Do you know how finalised this project is? I think it might look strange to cite an in prep paper that is not close to publication}
Furthermore, we hope it will be useful for researchers who are non-experts on primordial black holes, who need a reasonable PBH mass function to input into their pipelines for calculating e.g. observational constraints from pulsar timing arrays \cite{Afzal_2023,refId0} or microlensing experiments \cite{hsc-subaru,Mr_z_2024}. {The code could also provide a supporting tool for future and ongoing efforts toward a complete framework for computing gravitational wave signatures from PBHs, as is in progress within the LISA Cosmology Working Group.}

The current version of the code includes the three statistical methods briefly outlined in \cref{sec:stats} and assumes that the comoving curvature perturbation follows an approximately Gaussian distribution. We would like, in the future, to add the ability to deal with additional complications such as the softening of the equation of state during the QCD phase transition \cite{Byrnes_2018,Escriv__2023,Musco:2023dak} (relevant for $\sim M_\odot$ black holes) and non-Gaussianity \cite{Bullock:1996at, Ivanov:1997ia, Young:2013oia, Byrnes:2012yx, Tada:2015noa, Atal:2018neu, Franciolini:2018vbk,Kehagias:2019eil,Riccardi:2021rlf,Young:2022gfc,Gow:2022jfb,Matsubara_2022} as well as keeping the package up-to-date with theoretical advances as they are established. We welcome feedback and suggestions at \path{https://github.com/pipcole/pbhstat}.

%%%%%%%%%%%%%%%%%%%%%%%%%%%%%%%%%%%%%%%%%%

\begin{acknowledgments}
 The authors thank Jaume Garriga, Cristiano Germani, Andrew Gow and Ravi K. Sheth for helpful discussions. P.C. is supported by ERC Starting Grant No.~945155--GWmining, Cariplo Foundation Grant No.~2021-0555, MUR PRIN Grant No.~2022-Z9X4XS, MUR Grant ``Progetto Dipartimenti di Eccellenza 2023-2027'' (BiCoQ), and the ICSC National Research Centre funded by NextGenerationEU. The research of J.F. is supported by the
grant PID2022-136224NB-C22, funded by MCIN\allowbreak/\allowbreak AEI\allowbreak/10.13039\allowbreak/501100011033\allowbreak/\allowbreak FEDER,
UE, and by the grant\allowbreak/ 2021-SGR00872.
\end{acknowledgments}
\bibliographystyle{apsrev4-1}
\bibliography{main.bib}

\begin{thebibliography}{74}%
\makeatletter
\providecommand \@ifxundefined [1]{%
 \@ifx{#1\undefined}
}%
\providecommand \@ifnum [1]{%
 \ifnum #1\expandafter \@firstoftwo
 \else \expandafter \@secondoftwo
 \fi
}%
\providecommand \@ifx [1]{%
 \ifx #1\expandafter \@firstoftwo
 \else \expandafter \@secondoftwo
 \fi
}%
\providecommand \natexlab [1]{#1}%
\providecommand \enquote  [1]{``#1''}%
\providecommand \bibnamefont  [1]{#1}%
\providecommand \bibfnamefont [1]{#1}%
\providecommand \citenamefont [1]{#1}%
\providecommand \href@noop [0]{\@secondoftwo}%
\providecommand \href [0]{\begingroup \@sanitize@url \@href}%
\providecommand \@href[1]{\@@startlink{#1}\@@href}%
\providecommand \@@href[1]{\endgroup#1\@@endlink}%
\providecommand \@sanitize@url [0]{\catcode `\\12\catcode `\$12\catcode `\&12\catcode `\#12\catcode `\^12\catcode `\_12\catcode `\%12\relax}%
\providecommand \@@startlink[1]{}%
\providecommand \@@endlink[0]{}%
\providecommand \url  [0]{\begingroup\@sanitize@url \@url }%
\providecommand \@url [1]{\endgroup\@href {#1}{\urlprefix }}%
\providecommand \urlprefix  [0]{URL }%
\providecommand \Eprint [0]{\href }%
\providecommand \doibase [0]{http://dx.doi.org/}%
\providecommand \selectlanguage [0]{\@gobble}%
\providecommand \bibinfo  [0]{\@secondoftwo}%
\providecommand \bibfield  [0]{\@secondoftwo}%
\providecommand \translation [1]{[#1]}%
\providecommand \BibitemOpen [0]{}%
\providecommand \bibitemStop [0]{}%
\providecommand \bibitemNoStop [0]{.\EOS\space}%
\providecommand \EOS [0]{\spacefactor3000\relax}%
\providecommand \BibitemShut  [1]{\csname bibitem#1\endcsname}%
\let\auto@bib@innerbib\@empty
%</preamble>
\bibitem [{\citenamefont {Zel'dovich}\ and\ \citenamefont {Novikov}(1967)}]{Zeldovich:1967lct}%
  \BibitemOpen
  \bibfield  {author} {\bibinfo {author} {\bibfnamefont {Y.~B.}\ \bibnamefont {Zel'dovich}}\ and\ \bibinfo {author} {\bibfnamefont {I.~D.}\ \bibnamefont {Novikov}},\ }\href@noop {} {\bibfield  {journal} {\bibinfo  {journal} {Soviet Astron. AJ (Engl. Transl. ),}\ }\textbf {\bibinfo {volume} {10}},\ \bibinfo {pages} {602} (\bibinfo {year} {1967})}\BibitemShut {NoStop}%
\bibitem [{\citenamefont {Hawking}(1971)}]{Hawking:1971ei}%
  \BibitemOpen
  \bibfield  {author} {\bibinfo {author} {\bibfnamefont {S.}~\bibnamefont {Hawking}},\ }\href {\doibase 10.1093/mnras/152.1.75} {\bibfield  {journal} {\bibinfo  {journal} {Mon. Not. Roy. Astron. Soc.}\ }\textbf {\bibinfo {volume} {152}},\ \bibinfo {pages} {75} (\bibinfo {year} {1971})}\BibitemShut {NoStop}%
\bibitem [{\citenamefont {Carr}\ and\ \citenamefont {Hawking}(1974)}]{Carr:1974nx}%
  \BibitemOpen
  \bibfield  {author} {\bibinfo {author} {\bibfnamefont {B.~J.}\ \bibnamefont {Carr}}\ and\ \bibinfo {author} {\bibfnamefont {S.~W.}\ \bibnamefont {Hawking}},\ }\href {\doibase 10.1093/mnras/168.2.399} {\bibfield  {journal} {\bibinfo  {journal} {Mon. Not. Roy. Astron. Soc.}\ }\textbf {\bibinfo {volume} {168}},\ \bibinfo {pages} {399} (\bibinfo {year} {1974})}\BibitemShut {NoStop}%
\bibitem [{\citenamefont {Page}\ and\ \citenamefont {Hawking}(1976)}]{Page:1976df}%
  \BibitemOpen
  \bibfield  {author} {\bibinfo {author} {\bibfnamefont {D.~N.}\ \bibnamefont {Page}}\ and\ \bibinfo {author} {\bibfnamefont {S.~W.}\ \bibnamefont {Hawking}},\ }\href@noop {} {\bibfield  {journal} {\bibinfo  {journal} {Astrophys. J.}\ }\textbf {\bibinfo {volume} {206}},\ \bibinfo {pages} {1} (\bibinfo {year} {1976})}\BibitemShut {NoStop}%
\bibitem [{\citenamefont {MacGibbon}\ and\ \citenamefont {Carr}(1991)}]{MacGibbon:1991tj}%
  \BibitemOpen
  \bibfield  {author} {\bibinfo {author} {\bibfnamefont {J.~H.}\ \bibnamefont {MacGibbon}}\ and\ \bibinfo {author} {\bibfnamefont {B.~J.}\ \bibnamefont {Carr}},\ }\href@noop {} {\bibfield  {journal} {\bibinfo  {journal} {Astrophys. J.}\ }\textbf {\bibinfo {volume} {371}},\ \bibinfo {pages} {447} (\bibinfo {year} {1991})}\BibitemShut {NoStop}%
\bibitem [{\citenamefont {Barrau}\ \emph {et~al.}(2003)\citenamefont {Barrau}, \citenamefont {Boudoul}, \citenamefont {Donnard}, \citenamefont {Grain},\ and\ \citenamefont {Servant}}]{Barrau:2003nj}%
  \BibitemOpen
  \bibfield  {author} {\bibinfo {author} {\bibfnamefont {A.}~\bibnamefont {Barrau}}, \bibinfo {author} {\bibfnamefont {G.}~\bibnamefont {Boudoul}}, \bibinfo {author} {\bibfnamefont {J.}~\bibnamefont {Donnard}}, \bibinfo {author} {\bibfnamefont {J.}~\bibnamefont {Grain}}, \ and\ \bibinfo {author} {\bibfnamefont {G.}~\bibnamefont {Servant}},\ }\href {\doibase 10.1051/0004-6361:20021556} {\bibfield  {journal} {\bibinfo  {journal} {Astron. Astrophys.}\ }\textbf {\bibinfo {volume} {398}},\ \bibinfo {pages} {403} (\bibinfo {year} {2003})},\ \Eprint {http://arxiv.org/abs/astro-ph/0207395} {arXiv:astro-ph/0207395} \BibitemShut {NoStop}%
\bibitem [{\citenamefont {Sasaki}\ \emph {et~al.}(2016)\citenamefont {Sasaki}, \citenamefont {Suyama}, \citenamefont {Tanaka},\ and\ \citenamefont {Yokoyama}}]{Sasaki:2016jop}%
  \BibitemOpen
  \bibfield  {author} {\bibinfo {author} {\bibfnamefont {M.}~\bibnamefont {Sasaki}}, \bibinfo {author} {\bibfnamefont {T.}~\bibnamefont {Suyama}}, \bibinfo {author} {\bibfnamefont {T.}~\bibnamefont {Tanaka}}, \ and\ \bibinfo {author} {\bibfnamefont {S.}~\bibnamefont {Yokoyama}},\ }\href {\doibase 10.1103/PhysRevLett.117.061101} {\bibfield  {journal} {\bibinfo  {journal} {Phys. Rev. Lett.}\ }\textbf {\bibinfo {volume} {117}},\ \bibinfo {pages} {061101} (\bibinfo {year} {2016})},\ \Eprint {http://arxiv.org/abs/1603.08338} {arXiv:1603.08338} \BibitemShut {NoStop}%
\bibitem [{\citenamefont {Carr}\ \emph {et~al.}(2021)\citenamefont {Carr}, \citenamefont {Kohri}, \citenamefont {Sendouda},\ and\ \citenamefont {Yokoyama}}]{Carr:2020gox}%
  \BibitemOpen
  \bibfield  {author} {\bibinfo {author} {\bibfnamefont {B.}~\bibnamefont {Carr}}, \bibinfo {author} {\bibfnamefont {K.}~\bibnamefont {Kohri}}, \bibinfo {author} {\bibfnamefont {Y.}~\bibnamefont {Sendouda}}, \ and\ \bibinfo {author} {\bibfnamefont {J.}~\bibnamefont {Yokoyama}},\ }\href {\doibase 10.1088/1361-6633/ac1e31} {\bibfield  {journal} {\bibinfo  {journal} {Rept. Prog. Phys.}\ }\textbf {\bibinfo {volume} {84}},\ \bibinfo {pages} {116902} (\bibinfo {year} {2021})},\ \Eprint {http://arxiv.org/abs/2002.12778} {arXiv:2002.12778} \BibitemShut {NoStop}%
\bibitem [{\citenamefont {Green}\ and\ \citenamefont {Kavanagh}(2021)}]{Green:2020jor}%
  \BibitemOpen
  \bibfield  {author} {\bibinfo {author} {\bibfnamefont {A.~M.}\ \bibnamefont {Green}}\ and\ \bibinfo {author} {\bibfnamefont {B.~J.}\ \bibnamefont {Kavanagh}},\ }\href {\doibase 10.1088/1361-6471/abc534} {\bibfield  {journal} {\bibinfo  {journal} {J. Phys. G}\ }\textbf {\bibinfo {volume} {48}},\ \bibinfo {pages} {043001} (\bibinfo {year} {2021})},\ \Eprint {http://arxiv.org/abs/2007.10722} {arXiv:2007.10722} \BibitemShut {NoStop}%
\bibitem [{\citenamefont {Khlopov}(2010)}]{Khlopov_2010}%
  \BibitemOpen
  \bibfield  {author} {\bibinfo {author} {\bibfnamefont {M.~Y.}\ \bibnamefont {Khlopov}},\ }\href {\doibase 10.1088/1674-4527/10/6/001} {\bibfield  {journal} {\bibinfo  {journal} {Research in Astronomy and Astrophysics}\ }\textbf {\bibinfo {volume} {10}},\ \bibinfo {pages} {495–528} (\bibinfo {year} {2010})}\BibitemShut {NoStop}%
\bibitem [{\citenamefont {Ivanov}\ \emph {et~al.}(1994)\citenamefont {Ivanov}, \citenamefont {Naselsky},\ and\ \citenamefont {Novikov}}]{Ivanov:1994pa}%
  \BibitemOpen
  \bibfield  {author} {\bibinfo {author} {\bibfnamefont {P.}~\bibnamefont {Ivanov}}, \bibinfo {author} {\bibfnamefont {P.}~\bibnamefont {Naselsky}}, \ and\ \bibinfo {author} {\bibfnamefont {I.}~\bibnamefont {Novikov}},\ }\href {\doibase 10.1103/PhysRevD.50.7173} {\bibfield  {journal} {\bibinfo  {journal} {Phys. Rev. D}\ }\textbf {\bibinfo {volume} {50}},\ \bibinfo {pages} {7173} (\bibinfo {year} {1994})}\BibitemShut {NoStop}%
\bibitem [{\citenamefont {Carr}\ \emph {et~al.}(1994)\citenamefont {Carr}, \citenamefont {Gilbert},\ and\ \citenamefont {Lidsey}}]{Carr:1994ar}%
  \BibitemOpen
  \bibfield  {author} {\bibinfo {author} {\bibfnamefont {B.~J.}\ \bibnamefont {Carr}}, \bibinfo {author} {\bibfnamefont {J.~H.}\ \bibnamefont {Gilbert}}, \ and\ \bibinfo {author} {\bibfnamefont {J.~E.}\ \bibnamefont {Lidsey}},\ }\href {\doibase 10.1103/PhysRevD.50.4853} {\bibfield  {journal} {\bibinfo  {journal} {Phys. Rev. D}\ }\textbf {\bibinfo {volume} {50}},\ \bibinfo {pages} {4853} (\bibinfo {year} {1994})},\ \Eprint {http://arxiv.org/abs/astro-ph/9405027} {arXiv:astro-ph/9405027} \BibitemShut {NoStop}%
\bibitem [{\citenamefont {Garcia-Bellido}\ \emph {et~al.}(1996)\citenamefont {Garcia-Bellido}, \citenamefont {Linde},\ and\ \citenamefont {Wands}}]{Garcia-Bellido:1996mdl}%
  \BibitemOpen
  \bibfield  {author} {\bibinfo {author} {\bibfnamefont {J.}~\bibnamefont {Garcia-Bellido}}, \bibinfo {author} {\bibfnamefont {A.~D.}\ \bibnamefont {Linde}}, \ and\ \bibinfo {author} {\bibfnamefont {D.}~\bibnamefont {Wands}},\ }\href {\doibase 10.1103/PhysRevD.54.6040} {\bibfield  {journal} {\bibinfo  {journal} {Phys. Rev. D}\ }\textbf {\bibinfo {volume} {54}},\ \bibinfo {pages} {6040} (\bibinfo {year} {1996})},\ \Eprint {http://arxiv.org/abs/astro-ph/9605094} {arXiv:astro-ph/9605094} \BibitemShut {NoStop}%
\bibitem [{\citenamefont {Yoo}(2024)}]{yoo2024basicsprimordialblackhole}%
  \BibitemOpen
  \bibfield  {author} {\bibinfo {author} {\bibfnamefont {C.-M.}\ \bibnamefont {Yoo}},\ }\href {https://arxiv.org/abs/2211.13512} {\enquote {\bibinfo {title} {The basics of primordial black hole formation and abundance estimation},}\ } (\bibinfo {year} {2024}),\ \Eprint {http://arxiv.org/abs/2211.13512} {arXiv:2211.13512 [astro-ph.CO]} \BibitemShut {NoStop}%
\bibitem [{\citenamefont {Carr}\ and\ \citenamefont {Kuhnel}(2022)}]{Carr:2021bzv}%
  \BibitemOpen
  \bibfield  {author} {\bibinfo {author} {\bibfnamefont {B.}~\bibnamefont {Carr}}\ and\ \bibinfo {author} {\bibfnamefont {F.}~\bibnamefont {Kuhnel}},\ }\href {\doibase 10.21468/SciPostPhysLectNotes.48} {\bibfield  {journal} {\bibinfo  {journal} {SciPost Phys. Lect. Notes}\ }\textbf {\bibinfo {volume} {48}},\ \bibinfo {pages} {1} (\bibinfo {year} {2022})},\ \Eprint {http://arxiv.org/abs/2110.02821} {arXiv:2110.02821 [astro-ph.CO]} \BibitemShut {NoStop}%
\bibitem [{\citenamefont {Escriv{\`a}}\ \emph {et~al.}(2022)\citenamefont {Escriv{\`a}}, \citenamefont {Kuhnel},\ and\ \citenamefont {Tada}}]{Escriva:2022duf}%
  \BibitemOpen
  \bibfield  {author} {\bibinfo {author} {\bibfnamefont {A.}~\bibnamefont {Escriv{\`a}}}, \bibinfo {author} {\bibfnamefont {F.}~\bibnamefont {Kuhnel}}, \ and\ \bibinfo {author} {\bibfnamefont {Y.}~\bibnamefont {Tada}},\ }\href {\doibase 10.1016/B978-0-32-395636-9.00012-8} {\  (\bibinfo {year} {2022}),\ 10.1016/B978-0-32-395636-9.00012-8},\ \Eprint {http://arxiv.org/abs/2211.05767} {arXiv:2211.05767 [astro-ph.CO]} \BibitemShut {NoStop}%
\bibitem [{\citenamefont {Musco}\ \emph {et~al.}(2005)\citenamefont {Musco}, \citenamefont {Miller},\ and\ \citenamefont {Rezzolla}}]{Musco:2004ak}%
  \BibitemOpen
  \bibfield  {author} {\bibinfo {author} {\bibfnamefont {I.}~\bibnamefont {Musco}}, \bibinfo {author} {\bibfnamefont {J.~C.}\ \bibnamefont {Miller}}, \ and\ \bibinfo {author} {\bibfnamefont {L.}~\bibnamefont {Rezzolla}},\ }\href {\doibase 10.1088/0264-9381/22/7/013} {\bibfield  {journal} {\bibinfo  {journal} {Class. Quant. Grav.}\ }\textbf {\bibinfo {volume} {22}},\ \bibinfo {pages} {1405} (\bibinfo {year} {2005})},\ \Eprint {http://arxiv.org/abs/gr-qc/0412063} {arXiv:gr-qc/0412063} \BibitemShut {NoStop}%
\bibitem [{\citenamefont {Escriv{\`a}}(2020)}]{Escriva:2019nsa}%
  \BibitemOpen
  \bibfield  {author} {\bibinfo {author} {\bibfnamefont {A.}~\bibnamefont {Escriv{\`a}}},\ }\href {\doibase 10.1016/j.dark.2020.100466} {\bibfield  {journal} {\bibinfo  {journal} {Phys. Dark Univ.}\ }\textbf {\bibinfo {volume} {27}},\ \bibinfo {pages} {100466} (\bibinfo {year} {2020})},\ \Eprint {http://arxiv.org/abs/1907.13065} {arXiv:1907.13065 [gr-qc]} \BibitemShut {NoStop}%
\bibitem [{\citenamefont {Musco}(2025)}]{Musco:2025zft}%
  \BibitemOpen
  \bibfield  {author} {\bibinfo {author} {\bibfnamefont {I.}~\bibnamefont {Musco}},\ }\enquote {\bibinfo {title} {{Numerical Simulations of Primordial Black Holes}},}\ \ (\bibinfo {year} {2025})\BibitemShut {NoStop}%
\bibitem [{\citenamefont {Germani}\ and\ \citenamefont {Musco}(2019)}]{Germani:2018jgr}%
  \BibitemOpen
  \bibfield  {author} {\bibinfo {author} {\bibfnamefont {C.}~\bibnamefont {Germani}}\ and\ \bibinfo {author} {\bibfnamefont {I.}~\bibnamefont {Musco}},\ }\href {\doibase 10.1103/PhysRevLett.122.141302} {\bibfield  {journal} {\bibinfo  {journal} {Phys. Rev. Lett.}\ }\textbf {\bibinfo {volume} {122}},\ \bibinfo {pages} {141302} (\bibinfo {year} {2019})},\ \Eprint {http://arxiv.org/abs/1805.04087} {arXiv:1805.04087 [astro-ph.CO]} \BibitemShut {NoStop}%
\bibitem [{\citenamefont {Musco}(2019)}]{Musco:2018rwt}%
  \BibitemOpen
  \bibfield  {author} {\bibinfo {author} {\bibfnamefont {I.}~\bibnamefont {Musco}},\ }\href {\doibase 10.1103/PhysRevD.100.123524} {\bibfield  {journal} {\bibinfo  {journal} {Phys. Rev. D}\ }\textbf {\bibinfo {volume} {100}},\ \bibinfo {pages} {123524} (\bibinfo {year} {2019})},\ \Eprint {http://arxiv.org/abs/1809.02127} {arXiv:1809.02127 [gr-qc]} \BibitemShut {NoStop}%
\bibitem [{\citenamefont {Ando}\ \emph {et~al.}(2018)\citenamefont {Ando}, \citenamefont {Inomata},\ and\ \citenamefont {Kawasaki}}]{Ando_2018}%
  \BibitemOpen
  \bibfield  {author} {\bibinfo {author} {\bibfnamefont {K.}~\bibnamefont {Ando}}, \bibinfo {author} {\bibfnamefont {K.}~\bibnamefont {Inomata}}, \ and\ \bibinfo {author} {\bibfnamefont {M.}~\bibnamefont {Kawasaki}},\ }\href {\doibase 10.1103/physrevd.97.103528} {\bibfield  {journal} {\bibinfo  {journal} {Physical Review D}\ }\textbf {\bibinfo {volume} {97}} (\bibinfo {year} {2018}),\ 10.1103/physrevd.97.103528}\BibitemShut {NoStop}%
\bibitem [{\citenamefont {Yoo}\ \emph {et~al.}(2018)\citenamefont {Yoo}, \citenamefont {Harada}, \citenamefont {Garriga},\ and\ \citenamefont {Kohri}}]{Yoo_2018}%
  \BibitemOpen
  \bibfield  {author} {\bibinfo {author} {\bibfnamefont {C.-M.}\ \bibnamefont {Yoo}}, \bibinfo {author} {\bibfnamefont {T.}~\bibnamefont {Harada}}, \bibinfo {author} {\bibfnamefont {J.}~\bibnamefont {Garriga}}, \ and\ \bibinfo {author} {\bibfnamefont {K.}~\bibnamefont {Kohri}},\ }\href {\doibase 10.1093/ptep/pty120} {\bibfield  {journal} {\bibinfo  {journal} {Progress of Theoretical and Experimental Physics}\ }\textbf {\bibinfo {volume} {2018}} (\bibinfo {year} {2018}),\ 10.1093/ptep/pty120}\BibitemShut {NoStop}%
\bibitem [{\citenamefont {Young}(2025)}]{Young:2024jsu}%
  \BibitemOpen
  \bibfield  {author} {\bibinfo {author} {\bibfnamefont {S.}~\bibnamefont {Young}},\ }\enquote {\bibinfo {title} {{Computation of~the~Abundance of~Primordial Black Holes}},}\ \ (\bibinfo {year} {2025})\ \Eprint {http://arxiv.org/abs/2405.13259} {arXiv:2405.13259 [astro-ph.CO]} \BibitemShut {NoStop}%
\bibitem [{\citenamefont {Fumagalli}\ \emph {et~al.}(2025)\citenamefont {Fumagalli}, \citenamefont {Garriga}, \citenamefont {Germani},\ and\ \citenamefont {Sheth}}]{Fumagalli:2024kgg}%
  \BibitemOpen
  \bibfield  {author} {\bibinfo {author} {\bibfnamefont {J.}~\bibnamefont {Fumagalli}}, \bibinfo {author} {\bibfnamefont {J.}~\bibnamefont {Garriga}}, \bibinfo {author} {\bibfnamefont {C.}~\bibnamefont {Germani}}, \ and\ \bibinfo {author} {\bibfnamefont {R.~K.}\ \bibnamefont {Sheth}},\ }\href {\doibase 10.1103/k75n-3qz4} {\bibfield  {journal} {\bibinfo  {journal} {Phys. Rev. D}\ }\textbf {\bibinfo {volume} {111}},\ \bibinfo {pages} {123518} (\bibinfo {year} {2025})},\ \Eprint {http://arxiv.org/abs/2412.07709} {arXiv:2412.07709 [astro-ph.CO]} \BibitemShut {NoStop}%
\bibitem [{\citenamefont {Pi}\ \emph {et~al.}(2024)\citenamefont {Pi}, \citenamefont {Sasaki}, \citenamefont {Takhistov},\ and\ \citenamefont {Wang}}]{Pi:2024ert}%
  \BibitemOpen
  \bibfield  {author} {\bibinfo {author} {\bibfnamefont {S.}~\bibnamefont {Pi}}, \bibinfo {author} {\bibfnamefont {M.}~\bibnamefont {Sasaki}}, \bibinfo {author} {\bibfnamefont {V.}~\bibnamefont {Takhistov}}, \ and\ \bibinfo {author} {\bibfnamefont {J.}~\bibnamefont {Wang}},\ }\href@noop {} {\  (\bibinfo {year} {2024})},\ \Eprint {http://arxiv.org/abs/2501.00295} {arXiv:2501.00295 [astro-ph.CO]} \BibitemShut {NoStop}%
\bibitem [{\citenamefont {{Press}}\ and\ \citenamefont {{Schechter}}(1974)}]{press_original}%
  \BibitemOpen
  \bibfield  {author} {\bibinfo {author} {\bibfnamefont {W.~H.}\ \bibnamefont {{Press}}}\ and\ \bibinfo {author} {\bibfnamefont {P.}~\bibnamefont {{Schechter}}},\ }\href {\doibase 10.1086/152650} {\bibfield  {journal} {\bibinfo  {journal} {\apj}\ }\textbf {\bibinfo {volume} {187}},\ \bibinfo {pages} {425} (\bibinfo {year} {1974})}\BibitemShut {NoStop}%
\bibitem [{\citenamefont {Carr}(1975)}]{Carr:1975qj}%
  \BibitemOpen
  \bibfield  {author} {\bibinfo {author} {\bibfnamefont {B.~J.}\ \bibnamefont {Carr}},\ }\href {\doibase 10.1086/153853} {\bibfield  {journal} {\bibinfo  {journal} {Astrophys. J.}\ }\textbf {\bibinfo {volume} {201}},\ \bibinfo {pages} {1} (\bibinfo {year} {1975})}\BibitemShut {NoStop}%
\bibitem [{\citenamefont {{Bardeen}}\ \emph {et~al.}(1986)\citenamefont {{Bardeen}}, \citenamefont {{Bond}}, \citenamefont {{Kaiser}},\ and\ \citenamefont {{Szalay}}}]{peaks_original}%
  \BibitemOpen
  \bibfield  {author} {\bibinfo {author} {\bibfnamefont {J.~M.}\ \bibnamefont {{Bardeen}}}, \bibinfo {author} {\bibfnamefont {J.~R.}\ \bibnamefont {{Bond}}}, \bibinfo {author} {\bibfnamefont {N.}~\bibnamefont {{Kaiser}}}, \ and\ \bibinfo {author} {\bibfnamefont {A.~S.}\ \bibnamefont {{Szalay}}},\ }\href {\doibase 10.1086/164143} {\bibfield  {journal} {\bibinfo  {journal} {\apj}\ }\textbf {\bibinfo {volume} {304}},\ \bibinfo {pages} {15} (\bibinfo {year} {1986})}\BibitemShut {NoStop}%
\bibitem [{\citenamefont {Green}\ \emph {et~al.}(2004)\citenamefont {Green}, \citenamefont {Liddle}, \citenamefont {Malik},\ and\ \citenamefont {Sasaki}}]{Green:2004wb}%
  \BibitemOpen
  \bibfield  {author} {\bibinfo {author} {\bibfnamefont {A.~M.}\ \bibnamefont {Green}}, \bibinfo {author} {\bibfnamefont {A.~R.}\ \bibnamefont {Liddle}}, \bibinfo {author} {\bibfnamefont {K.~A.}\ \bibnamefont {Malik}}, \ and\ \bibinfo {author} {\bibfnamefont {M.}~\bibnamefont {Sasaki}},\ }\href {\doibase 10.1103/PhysRevD.70.041502} {\bibfield  {journal} {\bibinfo  {journal} {Phys. Rev. D}\ }\textbf {\bibinfo {volume} {70}},\ \bibinfo {pages} {041502} (\bibinfo {year} {2004})},\ \Eprint {http://arxiv.org/abs/astro-ph/0403181} {arXiv:astro-ph/0403181} \BibitemShut {NoStop}%
\bibitem [{\citenamefont {Young}\ \emph {et~al.}(2014)\citenamefont {Young}, \citenamefont {Byrnes},\ and\ \citenamefont {Sasaki}}]{Young_2014}%
  \BibitemOpen
  \bibfield  {author} {\bibinfo {author} {\bibfnamefont {S.}~\bibnamefont {Young}}, \bibinfo {author} {\bibfnamefont {C.~T.}\ \bibnamefont {Byrnes}}, \ and\ \bibinfo {author} {\bibfnamefont {M.}~\bibnamefont {Sasaki}},\ }\href {\doibase 10.1088/1475-7516/2014/07/045} {\bibfield  {journal} {\bibinfo  {journal} {Journal of Cosmology and Astroparticle Physics}\ }\textbf {\bibinfo {volume} {2014}},\ \bibinfo {pages} {045–045} (\bibinfo {year} {2014})}\BibitemShut {NoStop}%
\bibitem [{\citenamefont {Young}\ and\ \citenamefont {Musso}(2020)}]{Young:2020xmk}%
  \BibitemOpen
  \bibfield  {author} {\bibinfo {author} {\bibfnamefont {S.}~\bibnamefont {Young}}\ and\ \bibinfo {author} {\bibfnamefont {M.}~\bibnamefont {Musso}},\ }\href {\doibase 10.1088/1475-7516/2020/11/022} {\bibfield  {journal} {\bibinfo  {journal} {JCAP}\ }\textbf {\bibinfo {volume} {11}},\ \bibinfo {pages} {022} (\bibinfo {year} {2020})},\ \Eprint {http://arxiv.org/abs/2001.06469} {arXiv:2001.06469 [astro-ph.CO]} \BibitemShut {NoStop}%
\bibitem [{\citenamefont {Germani}\ and\ \citenamefont {Sheth}(2020)}]{Germani:2019zez}%
  \BibitemOpen
  \bibfield  {author} {\bibinfo {author} {\bibfnamefont {C.}~\bibnamefont {Germani}}\ and\ \bibinfo {author} {\bibfnamefont {R.~K.}\ \bibnamefont {Sheth}},\ }\href {\doibase 10.1103/PhysRevD.101.063520} {\bibfield  {journal} {\bibinfo  {journal} {Phys. Rev. D}\ }\textbf {\bibinfo {volume} {101}},\ \bibinfo {pages} {063520} (\bibinfo {year} {2020})},\ \Eprint {http://arxiv.org/abs/1912.07072} {arXiv:1912.07072 [astro-ph.CO]} \BibitemShut {NoStop}%
\bibitem [{\citenamefont {Germani}\ and\ \citenamefont {Sheth}(2023)}]{Germani:2023ojx}%
  \BibitemOpen
  \bibfield  {author} {\bibinfo {author} {\bibfnamefont {C.}~\bibnamefont {Germani}}\ and\ \bibinfo {author} {\bibfnamefont {R.~K.}\ \bibnamefont {Sheth}},\ }\href {\doibase 10.3390/universe9090421} {\bibfield  {journal} {\bibinfo  {journal} {Universe}\ }\textbf {\bibinfo {volume} {9}},\ \bibinfo {pages} {421} (\bibinfo {year} {2023})},\ \Eprint {http://arxiv.org/abs/2308.02971} {arXiv:2308.02971 [astro-ph.CO]} \BibitemShut {NoStop}%
\bibitem [{\citenamefont {Özsoy}\ and\ \citenamefont {Tasinato}(2023)}]{_zsoy_2023}%
  \BibitemOpen
  \bibfield  {author} {\bibinfo {author} {\bibfnamefont {O.}~\bibnamefont {Özsoy}}\ and\ \bibinfo {author} {\bibfnamefont {G.}~\bibnamefont {Tasinato}},\ }\href {\doibase 10.3390/universe9050203} {\bibfield  {journal} {\bibinfo  {journal} {Universe}\ }\textbf {\bibinfo {volume} {9}},\ \bibinfo {pages} {203} (\bibinfo {year} {2023})}\BibitemShut {NoStop}%
\bibitem [{\citenamefont {Pattison}\ \emph {et~al.}(2017)\citenamefont {Pattison}, \citenamefont {Vennin}, \citenamefont {Assadullahi},\ and\ \citenamefont {Wands}}]{Pattison_2017}%
  \BibitemOpen
  \bibfield  {author} {\bibinfo {author} {\bibfnamefont {C.}~\bibnamefont {Pattison}}, \bibinfo {author} {\bibfnamefont {V.}~\bibnamefont {Vennin}}, \bibinfo {author} {\bibfnamefont {H.}~\bibnamefont {Assadullahi}}, \ and\ \bibinfo {author} {\bibfnamefont {D.}~\bibnamefont {Wands}},\ }\href {\doibase 10.1088/1475-7516/2017/10/046} {\bibfield  {journal} {\bibinfo  {journal} {Journal of Cosmology and Astroparticle Physics}\ }\textbf {\bibinfo {volume} {2017}},\ \bibinfo {pages} {046–046} (\bibinfo {year} {2017})}\BibitemShut {NoStop}%
\bibitem [{\citenamefont {{Ezquiaga}}\ and\ \citenamefont {{Garc{\'\i}a-Bellido}}(2018)}]{2018JCAP...08..018E}%
  \BibitemOpen
  \bibfield  {author} {\bibinfo {author} {\bibfnamefont {J.~M.}\ \bibnamefont {{Ezquiaga}}}\ and\ \bibinfo {author} {\bibfnamefont {J.}~\bibnamefont {{Garc{\'\i}a-Bellido}}},\ }\href {\doibase 10.1088/1475-7516/2018/08/018} {\bibfield  {journal} {\bibinfo  {journal} {JCAP}\ }\textbf {\bibinfo {volume} {2018}},\ \bibinfo {eid} {018} (\bibinfo {year} {2018})},\ \Eprint {http://arxiv.org/abs/1805.06731} {arXiv:1805.06731 [astro-ph.CO]} \BibitemShut {NoStop}%
\bibitem [{\citenamefont {Biagetti}\ \emph {et~al.}(2021)\citenamefont {Biagetti}, \citenamefont {De~Luca}, \citenamefont {Franciolini}, \citenamefont {Kehagias},\ and\ \citenamefont {Riotto}}]{Biagetti:2021eep}%
  \BibitemOpen
  \bibfield  {author} {\bibinfo {author} {\bibfnamefont {M.}~\bibnamefont {Biagetti}}, \bibinfo {author} {\bibfnamefont {V.}~\bibnamefont {De~Luca}}, \bibinfo {author} {\bibfnamefont {G.}~\bibnamefont {Franciolini}}, \bibinfo {author} {\bibfnamefont {A.}~\bibnamefont {Kehagias}}, \ and\ \bibinfo {author} {\bibfnamefont {A.}~\bibnamefont {Riotto}},\ }\href {\doibase 10.1016/j.physletb.2021.136602} {\bibfield  {journal} {\bibinfo  {journal} {Phys. Lett. B}\ }\textbf {\bibinfo {volume} {820}},\ \bibinfo {pages} {136602} (\bibinfo {year} {2021})},\ \Eprint {http://arxiv.org/abs/2105.07810} {arXiv:2105.07810 [astro-ph.CO]} \BibitemShut {NoStop}%
\bibitem [{\citenamefont {Vennin}\ and\ \citenamefont {Wands}(2025)}]{Vennin_2025}%
  \BibitemOpen
  \bibfield  {author} {\bibinfo {author} {\bibfnamefont {V.}~\bibnamefont {Vennin}}\ and\ \bibinfo {author} {\bibfnamefont {D.}~\bibnamefont {Wands}},\ }\enquote {\bibinfo {title} {Quantum diffusion and large primordial perturbations from inflation},}\ in\ \href {\doibase 10.1007/978-981-97-8887-3_8} {\emph {\bibinfo {booktitle} {Primordial Black Holes}}}\ (\bibinfo  {publisher} {Springer Nature Singapore},\ \bibinfo {year} {2025})\ p.\ \bibinfo {pages} {201–227}\BibitemShut {NoStop}%
\bibitem [{\citenamefont {Bullock}\ and\ \citenamefont {Primack}(1997)}]{Bullock:1996at}%
  \BibitemOpen
  \bibfield  {author} {\bibinfo {author} {\bibfnamefont {J.~S.}\ \bibnamefont {Bullock}}\ and\ \bibinfo {author} {\bibfnamefont {J.~R.}\ \bibnamefont {Primack}},\ }\href {\doibase 10.1103/PhysRevD.55.7423} {\bibfield  {journal} {\bibinfo  {journal} {Phys. Rev. D}\ }\textbf {\bibinfo {volume} {55}},\ \bibinfo {pages} {7423} (\bibinfo {year} {1997})},\ \Eprint {http://arxiv.org/abs/astro-ph/9611106} {arXiv:astro-ph/9611106} \BibitemShut {NoStop}%
\bibitem [{\citenamefont {Ivanov}(1998)}]{Ivanov:1997ia}%
  \BibitemOpen
  \bibfield  {author} {\bibinfo {author} {\bibfnamefont {P.}~\bibnamefont {Ivanov}},\ }\href {\doibase 10.1103/PhysRevD.57.7145} {\bibfield  {journal} {\bibinfo  {journal} {Phys. Rev. D}\ }\textbf {\bibinfo {volume} {57}},\ \bibinfo {pages} {7145} (\bibinfo {year} {1998})},\ \Eprint {http://arxiv.org/abs/astro-ph/9708224} {arXiv:astro-ph/9708224} \BibitemShut {NoStop}%
\bibitem [{\citenamefont {Young}\ and\ \citenamefont {Byrnes}(2013)}]{Young:2013oia}%
  \BibitemOpen
  \bibfield  {author} {\bibinfo {author} {\bibfnamefont {S.}~\bibnamefont {Young}}\ and\ \bibinfo {author} {\bibfnamefont {C.~T.}\ \bibnamefont {Byrnes}},\ }\href {\doibase 10.1088/1475-7516/2013/08/052} {\bibfield  {journal} {\bibinfo  {journal} {JCAP}\ }\textbf {\bibinfo {volume} {08}},\ \bibinfo {pages} {052} (\bibinfo {year} {2013})},\ \Eprint {http://arxiv.org/abs/1307.4995} {arXiv:1307.4995} \BibitemShut {NoStop}%
\bibitem [{\citenamefont {Byrnes}\ \emph {et~al.}(2012)\citenamefont {Byrnes}, \citenamefont {Copeland},\ and\ \citenamefont {Green}}]{Byrnes:2012yx}%
  \BibitemOpen
  \bibfield  {author} {\bibinfo {author} {\bibfnamefont {C.~T.}\ \bibnamefont {Byrnes}}, \bibinfo {author} {\bibfnamefont {E.~J.}\ \bibnamefont {Copeland}}, \ and\ \bibinfo {author} {\bibfnamefont {A.~M.}\ \bibnamefont {Green}},\ }\href {\doibase 10.1103/PhysRevD.86.043512} {\bibfield  {journal} {\bibinfo  {journal} {Phys. Rev. D}\ }\textbf {\bibinfo {volume} {86}},\ \bibinfo {pages} {043512} (\bibinfo {year} {2012})},\ \Eprint {http://arxiv.org/abs/1206.4188} {arXiv:1206.4188} \BibitemShut {NoStop}%
\bibitem [{\citenamefont {Tada}\ and\ \citenamefont {Yokoyama}(2015)}]{Tada:2015noa}%
  \BibitemOpen
  \bibfield  {author} {\bibinfo {author} {\bibfnamefont {Y.}~\bibnamefont {Tada}}\ and\ \bibinfo {author} {\bibfnamefont {S.}~\bibnamefont {Yokoyama}},\ }\href {\doibase 10.1103/PhysRevD.91.123534} {\bibfield  {journal} {\bibinfo  {journal} {Phys. Rev. D}\ }\textbf {\bibinfo {volume} {91}},\ \bibinfo {pages} {123534} (\bibinfo {year} {2015})},\ \Eprint {http://arxiv.org/abs/1502.01124} {arXiv:1502.01124} \BibitemShut {NoStop}%
\bibitem [{\citenamefont {Atal}\ and\ \citenamefont {Germani}(2019)}]{Atal:2018neu}%
  \BibitemOpen
  \bibfield  {author} {\bibinfo {author} {\bibfnamefont {V.}~\bibnamefont {Atal}}\ and\ \bibinfo {author} {\bibfnamefont {C.}~\bibnamefont {Germani}},\ }\href {\doibase 10.1016/j.dark.2019.100275} {\bibfield  {journal} {\bibinfo  {journal} {Phys. Dark Univ.}\ }\textbf {\bibinfo {volume} {24}},\ \bibinfo {pages} {100275} (\bibinfo {year} {2019})},\ \Eprint {http://arxiv.org/abs/1811.07857} {arXiv:1811.07857} \BibitemShut {NoStop}%
\bibitem [{\citenamefont {Franciolini}\ \emph {et~al.}(2018)\citenamefont {Franciolini}, \citenamefont {Kehagias}, \citenamefont {Matarrese},\ and\ \citenamefont {Riotto}}]{Franciolini:2018vbk}%
  \BibitemOpen
  \bibfield  {author} {\bibinfo {author} {\bibfnamefont {G.}~\bibnamefont {Franciolini}}, \bibinfo {author} {\bibfnamefont {A.}~\bibnamefont {Kehagias}}, \bibinfo {author} {\bibfnamefont {S.}~\bibnamefont {Matarrese}}, \ and\ \bibinfo {author} {\bibfnamefont {A.}~\bibnamefont {Riotto}},\ }\href {\doibase 10.1088/1475-7516/2018/03/016} {\bibfield  {journal} {\bibinfo  {journal} {JCAP}\ }\textbf {\bibinfo {volume} {03}},\ \bibinfo {pages} {016} (\bibinfo {year} {2018})},\ \Eprint {http://arxiv.org/abs/1801.09415} {arXiv:1801.09415} \BibitemShut {NoStop}%
\bibitem [{\citenamefont {Kehagias}\ \emph {et~al.}(2019)\citenamefont {Kehagias}, \citenamefont {Musco},\ and\ \citenamefont {Riotto}}]{Kehagias:2019eil}%
  \BibitemOpen
  \bibfield  {author} {\bibinfo {author} {\bibfnamefont {A.}~\bibnamefont {Kehagias}}, \bibinfo {author} {\bibfnamefont {I.}~\bibnamefont {Musco}}, \ and\ \bibinfo {author} {\bibfnamefont {A.}~\bibnamefont {Riotto}},\ }\href {\doibase 10.1088/1475-7516/2019/12/029} {\bibfield  {journal} {\bibinfo  {journal} {JCAP}\ }\textbf {\bibinfo {volume} {12}},\ \bibinfo {pages} {029} (\bibinfo {year} {2019})},\ \Eprint {http://arxiv.org/abs/1906.07135} {arXiv:1906.07135 [astro-ph.CO]} \BibitemShut {NoStop}%
\bibitem [{\citenamefont {Riccardi}\ \emph {et~al.}(2021)\citenamefont {Riccardi}, \citenamefont {Taoso},\ and\ \citenamefont {Urbano}}]{Riccardi:2021rlf}%
  \BibitemOpen
  \bibfield  {author} {\bibinfo {author} {\bibfnamefont {F.}~\bibnamefont {Riccardi}}, \bibinfo {author} {\bibfnamefont {M.}~\bibnamefont {Taoso}}, \ and\ \bibinfo {author} {\bibfnamefont {A.}~\bibnamefont {Urbano}},\ }\href {\doibase 10.1088/1475-7516/2021/08/060} {\bibfield  {journal} {\bibinfo  {journal} {JCAP}\ }\textbf {\bibinfo {volume} {08}},\ \bibinfo {pages} {060} (\bibinfo {year} {2021})},\ \Eprint {http://arxiv.org/abs/2102.04084} {arXiv:2102.04084 [astro-ph.CO]} \BibitemShut {NoStop}%
\bibitem [{\citenamefont {Gow}\ \emph {et~al.}(2023)\citenamefont {Gow}, \citenamefont {Assadullahi}, \citenamefont {Jackson}, \citenamefont {Koyama}, \citenamefont {Vennin},\ and\ \citenamefont {Wands}}]{Gow:2022jfb}%
  \BibitemOpen
  \bibfield  {author} {\bibinfo {author} {\bibfnamefont {A.~D.}\ \bibnamefont {Gow}}, \bibinfo {author} {\bibfnamefont {H.}~\bibnamefont {Assadullahi}}, \bibinfo {author} {\bibfnamefont {J.~H.~P.}\ \bibnamefont {Jackson}}, \bibinfo {author} {\bibfnamefont {K.}~\bibnamefont {Koyama}}, \bibinfo {author} {\bibfnamefont {V.}~\bibnamefont {Vennin}}, \ and\ \bibinfo {author} {\bibfnamefont {D.}~\bibnamefont {Wands}},\ }\href {\doibase 10.1209/0295-5075/acd417} {\bibfield  {journal} {\bibinfo  {journal} {EPL}\ }\textbf {\bibinfo {volume} {142}},\ \bibinfo {pages} {49001} (\bibinfo {year} {2023})},\ \Eprint {http://arxiv.org/abs/2211.08348} {arXiv:2211.08348 [astro-ph.CO]} \BibitemShut {NoStop}%
\bibitem [{\citenamefont {Matsubara}\ and\ \citenamefont {Sasaki}(2022)}]{Matsubara_2022}%
  \BibitemOpen
  \bibfield  {author} {\bibinfo {author} {\bibfnamefont {T.}~\bibnamefont {Matsubara}}\ and\ \bibinfo {author} {\bibfnamefont {M.}~\bibnamefont {Sasaki}},\ }\href {\doibase 10.1088/1475-7516/2022/10/094} {\bibfield  {journal} {\bibinfo  {journal} {Journal of Cosmology and Astroparticle Physics}\ }\textbf {\bibinfo {volume} {2022}},\ \bibinfo {pages} {094} (\bibinfo {year} {2022})}\BibitemShut {NoStop}%
\bibitem [{\citenamefont {Nakama}\ \emph {et~al.}(2017)\citenamefont {Nakama}, \citenamefont {Silk},\ and\ \citenamefont {Kamionkowski}}]{Nakama_2017}%
  \BibitemOpen
  \bibfield  {author} {\bibinfo {author} {\bibfnamefont {T.}~\bibnamefont {Nakama}}, \bibinfo {author} {\bibfnamefont {J.}~\bibnamefont {Silk}}, \ and\ \bibinfo {author} {\bibfnamefont {M.}~\bibnamefont {Kamionkowski}},\ }\href {\doibase 10.1103/physrevd.95.043511} {\bibfield  {journal} {\bibinfo  {journal} {Physical Review D}\ }\textbf {\bibinfo {volume} {95}} (\bibinfo {year} {2017}),\ 10.1103/physrevd.95.043511}\BibitemShut {NoStop}%
\bibitem [{\citenamefont {Shibata}\ and\ \citenamefont {Sasaki}(1999)}]{Shibata:1999zs}%
  \BibitemOpen
  \bibfield  {author} {\bibinfo {author} {\bibfnamefont {M.}~\bibnamefont {Shibata}}\ and\ \bibinfo {author} {\bibfnamefont {M.}~\bibnamefont {Sasaki}},\ }\href {\doibase 10.1103/PhysRevD.60.084002} {\bibfield  {journal} {\bibinfo  {journal} {Phys. Rev. D}\ }\textbf {\bibinfo {volume} {60}},\ \bibinfo {pages} {084002} (\bibinfo {year} {1999})},\ \Eprint {http://arxiv.org/abs/gr-qc/9905064} {arXiv:gr-qc/9905064} \BibitemShut {NoStop}%
\bibitem [{\citenamefont {Harada}\ \emph {et~al.}(2015)\citenamefont {Harada}, \citenamefont {Yoo}, \citenamefont {Nakama},\ and\ \citenamefont {Koga}}]{Harada:2015yda}%
  \BibitemOpen
  \bibfield  {author} {\bibinfo {author} {\bibfnamefont {T.}~\bibnamefont {Harada}}, \bibinfo {author} {\bibfnamefont {C.-M.}\ \bibnamefont {Yoo}}, \bibinfo {author} {\bibfnamefont {T.}~\bibnamefont {Nakama}}, \ and\ \bibinfo {author} {\bibfnamefont {Y.}~\bibnamefont {Koga}},\ }\href {\doibase 10.1103/PhysRevD.91.084057} {\bibfield  {journal} {\bibinfo  {journal} {Phys. Rev. D}\ }\textbf {\bibinfo {volume} {91}},\ \bibinfo {pages} {084057} (\bibinfo {year} {2015})},\ \Eprint {http://arxiv.org/abs/1503.03934} {arXiv:1503.03934 [gr-qc]} \BibitemShut {NoStop}%
\bibitem [{\citenamefont {Uehara}\ \emph {et~al.}(2025)\citenamefont {Uehara}, \citenamefont {Escriv{\`a}}, \citenamefont {Harada}, \citenamefont {Saito},\ and\ \citenamefont {Yoo}}]{Uehara:2024yyp}%
  \BibitemOpen
  \bibfield  {author} {\bibinfo {author} {\bibfnamefont {K.}~\bibnamefont {Uehara}}, \bibinfo {author} {\bibfnamefont {A.}~\bibnamefont {Escriv{\`a}}}, \bibinfo {author} {\bibfnamefont {T.}~\bibnamefont {Harada}}, \bibinfo {author} {\bibfnamefont {D.}~\bibnamefont {Saito}}, \ and\ \bibinfo {author} {\bibfnamefont {C.-M.}\ \bibnamefont {Yoo}},\ }\href {\doibase 10.1088/1475-7516/2025/01/003} {\bibfield  {journal} {\bibinfo  {journal} {JCAP}\ }\textbf {\bibinfo {volume} {01}},\ \bibinfo {pages} {003} (\bibinfo {year} {2025})},\ \Eprint {http://arxiv.org/abs/2401.06329} {arXiv:2401.06329 [gr-qc]} \BibitemShut {NoStop}%
\bibitem [{\citenamefont {Shimada}\ \emph {et~al.}(2025)\citenamefont {Shimada}, \citenamefont {Escriv{\'a}}, \citenamefont {Saito}, \citenamefont {Uehara},\ and\ \citenamefont {Yoo}}]{Shimada:2024eec}%
  \BibitemOpen
  \bibfield  {author} {\bibinfo {author} {\bibfnamefont {M.}~\bibnamefont {Shimada}}, \bibinfo {author} {\bibfnamefont {A.}~\bibnamefont {Escriv{\'a}}}, \bibinfo {author} {\bibfnamefont {D.}~\bibnamefont {Saito}}, \bibinfo {author} {\bibfnamefont {K.}~\bibnamefont {Uehara}}, \ and\ \bibinfo {author} {\bibfnamefont {C.-M.}\ \bibnamefont {Yoo}},\ }\href {\doibase 10.1088/1475-7516/2025/02/018} {\bibfield  {journal} {\bibinfo  {journal} {JCAP}\ }\textbf {\bibinfo {volume} {02}},\ \bibinfo {pages} {018} (\bibinfo {year} {2025})},\ \Eprint {http://arxiv.org/abs/2411.07648} {arXiv:2411.07648 [gr-qc]} \BibitemShut {NoStop}%
\bibitem [{\citenamefont {Inui}\ \emph {et~al.}(2025)\citenamefont {Inui}, \citenamefont {Joana}, \citenamefont {Motohashi}, \citenamefont {Pi}, \citenamefont {Tada},\ and\ \citenamefont {Yokoyama}}]{Inui:2024fgk}%
  \BibitemOpen
  \bibfield  {author} {\bibinfo {author} {\bibfnamefont {R.}~\bibnamefont {Inui}}, \bibinfo {author} {\bibfnamefont {C.}~\bibnamefont {Joana}}, \bibinfo {author} {\bibfnamefont {H.}~\bibnamefont {Motohashi}}, \bibinfo {author} {\bibfnamefont {S.}~\bibnamefont {Pi}}, \bibinfo {author} {\bibfnamefont {Y.}~\bibnamefont {Tada}}, \ and\ \bibinfo {author} {\bibfnamefont {S.}~\bibnamefont {Yokoyama}},\ }\href {\doibase 10.1088/1475-7516/2025/03/021} {\bibfield  {journal} {\bibinfo  {journal} {JCAP}\ }\textbf {\bibinfo {volume} {03}},\ \bibinfo {pages} {021} (\bibinfo {year} {2025})},\ \Eprint {http://arxiv.org/abs/2411.07647} {arXiv:2411.07647 [astro-ph.CO]} \BibitemShut {NoStop}%
\bibitem [{\citenamefont {Escriv{\`a}}(2025)}]{Escriva:2025rja}%
  \BibitemOpen
  \bibfield  {author} {\bibinfo {author} {\bibfnamefont {A.}~\bibnamefont {Escriv{\`a}}},\ }\href@noop {} {\  (\bibinfo {year} {2025})},\ \Eprint {http://arxiv.org/abs/2504.05814} {arXiv:2504.05814 [astro-ph.CO]} \BibitemShut {NoStop}%
\bibitem [{\citenamefont {Gow}\ \emph {et~al.}(2021)\citenamefont {Gow}, \citenamefont {Byrnes}, \citenamefont {Cole},\ and\ \citenamefont {Young}}]{Gow:2020bzo}%
  \BibitemOpen
  \bibfield  {author} {\bibinfo {author} {\bibfnamefont {A.~D.}\ \bibnamefont {Gow}}, \bibinfo {author} {\bibfnamefont {C.~T.}\ \bibnamefont {Byrnes}}, \bibinfo {author} {\bibfnamefont {P.~S.}\ \bibnamefont {Cole}}, \ and\ \bibinfo {author} {\bibfnamefont {S.}~\bibnamefont {Young}},\ }\href {\doibase 10.1088/1475-7516/2021/02/002} {\bibfield  {journal} {\bibinfo  {journal} {JCAP}\ }\textbf {\bibinfo {volume} {02}},\ \bibinfo {pages} {002} (\bibinfo {year} {2021})},\ \Eprint {http://arxiv.org/abs/2008.03289} {arXiv:2008.03289 [astro-ph.CO]} \BibitemShut {NoStop}%
\bibitem [{\citenamefont {Young}(2019)}]{Young:2019osy}%
  \BibitemOpen
  \bibfield  {author} {\bibinfo {author} {\bibfnamefont {S.}~\bibnamefont {Young}},\ }\href {\doibase 10.1142/S0218271820300025} {\bibfield  {journal} {\bibinfo  {journal} {Int. J. Mod. Phys. D}\ }\textbf {\bibinfo {volume} {29}},\ \bibinfo {pages} {2030002} (\bibinfo {year} {2019})},\ \Eprint {http://arxiv.org/abs/1905.01230} {arXiv:1905.01230 [astro-ph.CO]} \BibitemShut {NoStop}%
\bibitem [{\citenamefont {Niemeyer}\ and\ \citenamefont {Jedamzik}(1998)}]{PhysRevLett.80.5481}%
  \BibitemOpen
  \bibfield  {author} {\bibinfo {author} {\bibfnamefont {J.~C.}\ \bibnamefont {Niemeyer}}\ and\ \bibinfo {author} {\bibfnamefont {K.}~\bibnamefont {Jedamzik}},\ }\href {\doibase 10.1103/PhysRevLett.80.5481} {\bibfield  {journal} {\bibinfo  {journal} {Phys. Rev. Lett.}\ }\textbf {\bibinfo {volume} {80}},\ \bibinfo {pages} {5481} (\bibinfo {year} {1998})}\BibitemShut {NoStop}%
\bibitem [{\citenamefont {Young}\ \emph {et~al.}(2019)\citenamefont {Young}, \citenamefont {Musco},\ and\ \citenamefont {Byrnes}}]{Young:2019yug}%
  \BibitemOpen
  \bibfield  {author} {\bibinfo {author} {\bibfnamefont {S.}~\bibnamefont {Young}}, \bibinfo {author} {\bibfnamefont {I.}~\bibnamefont {Musco}}, \ and\ \bibinfo {author} {\bibfnamefont {C.~T.}\ \bibnamefont {Byrnes}},\ }\href {\doibase 10.1088/1475-7516/2019/11/012} {\bibfield  {journal} {\bibinfo  {journal} {JCAP}\ }\textbf {\bibinfo {volume} {11}},\ \bibinfo {pages} {012} (\bibinfo {year} {2019})},\ \Eprint {http://arxiv.org/abs/1904.00984} {arXiv:1904.00984 [astro-ph.CO]} \BibitemShut {NoStop}%
\bibitem [{\citenamefont {Escriv{\`a}}\ \emph {et~al.}(2020)\citenamefont {Escriv{\`a}}, \citenamefont {Germani},\ and\ \citenamefont {Sheth}}]{Escriva:2019phb}%
  \BibitemOpen
  \bibfield  {author} {\bibinfo {author} {\bibfnamefont {A.}~\bibnamefont {Escriv{\`a}}}, \bibinfo {author} {\bibfnamefont {C.}~\bibnamefont {Germani}}, \ and\ \bibinfo {author} {\bibfnamefont {R.~K.}\ \bibnamefont {Sheth}},\ }\href {\doibase 10.1103/PhysRevD.101.044022} {\bibfield  {journal} {\bibinfo  {journal} {Phys. Rev. D}\ }\textbf {\bibinfo {volume} {101}},\ \bibinfo {pages} {044022} (\bibinfo {year} {2020})},\ \Eprint {http://arxiv.org/abs/1907.13311} {arXiv:1907.13311 [gr-qc]} \BibitemShut {NoStop}%
\bibitem [{\citenamefont {Kehagias}\ \emph {et~al.}(2025)\citenamefont {Kehagias}, \citenamefont {Perrone},\ and\ \citenamefont {Riotto}}]{Kehagias:2024kgk}%
  \BibitemOpen
  \bibfield  {author} {\bibinfo {author} {\bibfnamefont {A.}~\bibnamefont {Kehagias}}, \bibinfo {author} {\bibfnamefont {D.}~\bibnamefont {Perrone}}, \ and\ \bibinfo {author} {\bibfnamefont {A.}~\bibnamefont {Riotto}},\ }\href {\doibase 10.1088/1361-6382/adaffc} {\bibfield  {journal} {\bibinfo  {journal} {Class. Quant. Grav.}\ }\textbf {\bibinfo {volume} {42}},\ \bibinfo {pages} {055010} (\bibinfo {year} {2025})},\ \Eprint {http://arxiv.org/abs/2405.05208} {arXiv:2405.05208 [astro-ph.CO]} \BibitemShut {NoStop}%
\bibitem [{\citenamefont {Wang}\ \emph {et~al.}(2021)\citenamefont {Wang}, \citenamefont {Liu}, \citenamefont {Su},\ and\ \citenamefont {Li}}]{PhysRevD.104.083546}%
  \BibitemOpen
  \bibfield  {author} {\bibinfo {author} {\bibfnamefont {Q.}~\bibnamefont {Wang}}, \bibinfo {author} {\bibfnamefont {Y.-C.}\ \bibnamefont {Liu}}, \bibinfo {author} {\bibfnamefont {B.-Y.}\ \bibnamefont {Su}}, \ and\ \bibinfo {author} {\bibfnamefont {N.}~\bibnamefont {Li}},\ }\href {\doibase 10.1103/PhysRevD.104.083546} {\bibfield  {journal} {\bibinfo  {journal} {Phys. Rev. D}\ }\textbf {\bibinfo {volume} {104}},\ \bibinfo {pages} {083546} (\bibinfo {year} {2021})}\BibitemShut {NoStop}%
\bibitem [{\citenamefont {Kavanagh}(2019)}]{bradley_j_kavanagh_2019_3538999}%
  \BibitemOpen
  \bibfield  {author} {\bibinfo {author} {\bibfnamefont {B.~J.}\ \bibnamefont {Kavanagh}},\ }\href {\doibase 10.5281/zenodo.3538999} {\enquote {\bibinfo {title} {bradkav/pbhbounds: Release version},}\ } (\bibinfo {year} {2019})\BibitemShut {NoStop}%
\bibitem [{\citenamefont {Bellomo}\ \emph {et~al.}(2018)\citenamefont {Bellomo}, \citenamefont {Bernal}, \citenamefont {Raccanelli},\ and\ \citenamefont {Verde}}]{Bellomo:2017zsr}%
  \BibitemOpen
  \bibfield  {author} {\bibinfo {author} {\bibfnamefont {N.}~\bibnamefont {Bellomo}}, \bibinfo {author} {\bibfnamefont {J.~L.}\ \bibnamefont {Bernal}}, \bibinfo {author} {\bibfnamefont {A.}~\bibnamefont {Raccanelli}}, \ and\ \bibinfo {author} {\bibfnamefont {L.}~\bibnamefont {Verde}},\ }\href {\doibase 10.1088/1475-7516/2018/01/004} {\bibfield  {journal} {\bibinfo  {journal} {JCAP}\ }\textbf {\bibinfo {volume} {01}},\ \bibinfo {pages} {004} (\bibinfo {year} {2018})},\ \Eprint {http://arxiv.org/abs/1709.07467} {arXiv:1709.07467 [astro-ph.CO]} \BibitemShut {NoStop}%
\bibitem [{\citenamefont {Afzal}(2023)}]{Afzal_2023}%
  \BibitemOpen
  \bibfield  {author} {\bibinfo {author} {\bibfnamefont {A.~e.~a.}\ \bibnamefont {Afzal}},\ }\href {\doibase 10.3847/2041-8213/acdc91} {\bibfield  {journal} {\bibinfo  {journal} {The Astrophysical Journal Letters}\ }\textbf {\bibinfo {volume} {951}},\ \bibinfo {pages} {L11} (\bibinfo {year} {2023})}\BibitemShut {NoStop}%
\bibitem [{\citenamefont {et~al.}(2024)}]{refId0}%
  \BibitemOpen
  \bibfield  {author} {\bibinfo {author} {\bibfnamefont {E.}~\bibnamefont {et~al.}},\ }\href {\doibase 10.1051/0004-6361/202347433} {\bibfield  {journal} {\bibinfo  {journal} {A\&A}\ }\textbf {\bibinfo {volume} {685}},\ \bibinfo {pages} {A94} (\bibinfo {year} {2024})}\BibitemShut {NoStop}%
\bibitem [{\citenamefont {Niikura}\ \emph {et~al.}(2019)\citenamefont {Niikura}, \citenamefont {Takada}, \citenamefont {Yasuda}, \citenamefont {Lupton}, \citenamefont {Sumi}, \citenamefont {More}, \citenamefont {Kurita}, \citenamefont {Sugiyama}, \citenamefont {More}, \citenamefont {Oguri},\ and\ \citenamefont {Chiba}}]{hsc-subaru}%
  \BibitemOpen
  \bibfield  {author} {\bibinfo {author} {\bibfnamefont {H.}~\bibnamefont {Niikura}}, \bibinfo {author} {\bibfnamefont {M.}~\bibnamefont {Takada}}, \bibinfo {author} {\bibfnamefont {N.}~\bibnamefont {Yasuda}}, \bibinfo {author} {\bibfnamefont {R.~H.}\ \bibnamefont {Lupton}}, \bibinfo {author} {\bibfnamefont {T.}~\bibnamefont {Sumi}}, \bibinfo {author} {\bibfnamefont {S.}~\bibnamefont {More}}, \bibinfo {author} {\bibfnamefont {T.}~\bibnamefont {Kurita}}, \bibinfo {author} {\bibfnamefont {S.}~\bibnamefont {Sugiyama}}, \bibinfo {author} {\bibfnamefont {A.}~\bibnamefont {More}}, \bibinfo {author} {\bibfnamefont {M.}~\bibnamefont {Oguri}}, \ and\ \bibinfo {author} {\bibfnamefont {M.}~\bibnamefont {Chiba}},\ }\href {\doibase 10.1038/s41550-019-0723-1} {\bibfield  {journal} {\bibinfo  {journal} {Nature Astronomy}\ }\textbf {\bibinfo {volume} {3}},\ \bibinfo {pages} {524} (\bibinfo {year} {2019})}\BibitemShut {NoStop}%
\bibitem [{\citenamefont {Mróz}\ \emph {et~al.}(2024)\citenamefont {Mróz}, \citenamefont {Udalski}, \citenamefont {Szymański}, \citenamefont {Soszyński}, \citenamefont {Pietrukowicz}, \citenamefont {Kozłowski}, \citenamefont {Poleski}, \citenamefont {Skowron}, \citenamefont {Ulaczyk}, \citenamefont {Gromadzki}, \citenamefont {Rybicki}, \citenamefont {Iwanek}, \citenamefont {Wrona},\ and\ \citenamefont {Mróz}}]{Mr_z_2024}%
  \BibitemOpen
  \bibfield  {author} {\bibinfo {author} {\bibfnamefont {P.}~\bibnamefont {Mróz}}, \bibinfo {author} {\bibfnamefont {A.}~\bibnamefont {Udalski}}, \bibinfo {author} {\bibfnamefont {M.~K.}\ \bibnamefont {Szymański}}, \bibinfo {author} {\bibfnamefont {I.}~\bibnamefont {Soszyński}}, \bibinfo {author} {\bibfnamefont {P.}~\bibnamefont {Pietrukowicz}}, \bibinfo {author} {\bibfnamefont {S.}~\bibnamefont {Kozłowski}}, \bibinfo {author} {\bibfnamefont {R.}~\bibnamefont {Poleski}}, \bibinfo {author} {\bibfnamefont {J.}~\bibnamefont {Skowron}}, \bibinfo {author} {\bibfnamefont {K.}~\bibnamefont {Ulaczyk}}, \bibinfo {author} {\bibfnamefont {M.}~\bibnamefont {Gromadzki}}, \bibinfo {author} {\bibfnamefont {K.}~\bibnamefont {Rybicki}}, \bibinfo {author} {\bibfnamefont {P.}~\bibnamefont {Iwanek}}, \bibinfo {author} {\bibfnamefont {M.}~\bibnamefont {Wrona}}, \ and\ \bibinfo {author} {\bibfnamefont {M.~J.}\ \bibnamefont {Mróz}},\ }\href {\doibase 10.3847/2041-8213/ad8e68} {\bibfield  {journal} {\bibinfo  {journal} {The
  Astrophysical Journal Letters}\ }\textbf {\bibinfo {volume} {976}},\ \bibinfo {pages} {L19} (\bibinfo {year} {2024})}\BibitemShut {NoStop}%
\bibitem [{\citenamefont {Byrnes}\ \emph {et~al.}(2018)\citenamefont {Byrnes}, \citenamefont {Hindmarsh}, \citenamefont {Young},\ and\ \citenamefont {Hawkins}}]{Byrnes_2018}%
  \BibitemOpen
  \bibfield  {author} {\bibinfo {author} {\bibfnamefont {C.~T.}\ \bibnamefont {Byrnes}}, \bibinfo {author} {\bibfnamefont {M.}~\bibnamefont {Hindmarsh}}, \bibinfo {author} {\bibfnamefont {S.}~\bibnamefont {Young}}, \ and\ \bibinfo {author} {\bibfnamefont {M.~R.}\ \bibnamefont {Hawkins}},\ }\href {\doibase 10.1088/1475-7516/2018/08/041} {\bibfield  {journal} {\bibinfo  {journal} {Journal of Cosmology and Astroparticle Physics}\ }\textbf {\bibinfo {volume} {2018}},\ \bibinfo {pages} {041–041} (\bibinfo {year} {2018})}\BibitemShut {NoStop}%
\bibitem [{\citenamefont {Escrivà}\ \emph {et~al.}(2023)\citenamefont {Escrivà}, \citenamefont {Bagui},\ and\ \citenamefont {Clesse}}]{Escriv__2023}%
  \BibitemOpen
  \bibfield  {author} {\bibinfo {author} {\bibfnamefont {A.}~\bibnamefont {Escrivà}}, \bibinfo {author} {\bibfnamefont {E.}~\bibnamefont {Bagui}}, \ and\ \bibinfo {author} {\bibfnamefont {S.}~\bibnamefont {Clesse}},\ }\href {\doibase 10.1088/1475-7516/2023/05/004} {\bibfield  {journal} {\bibinfo  {journal} {Journal of Cosmology and Astroparticle Physics}\ }\textbf {\bibinfo {volume} {2023}},\ \bibinfo {pages} {004} (\bibinfo {year} {2023})}\BibitemShut {NoStop}%
\bibitem [{\citenamefont {Musco}\ \emph {et~al.}(2024)\citenamefont {Musco}, \citenamefont {Jedamzik},\ and\ \citenamefont {Young}}]{Musco:2023dak}%
  \BibitemOpen
  \bibfield  {author} {\bibinfo {author} {\bibfnamefont {I.}~\bibnamefont {Musco}}, \bibinfo {author} {\bibfnamefont {K.}~\bibnamefont {Jedamzik}}, \ and\ \bibinfo {author} {\bibfnamefont {S.}~\bibnamefont {Young}},\ }\href {\doibase 10.1103/PhysRevD.109.083506} {\bibfield  {journal} {\bibinfo  {journal} {Phys. Rev. D}\ }\textbf {\bibinfo {volume} {109}},\ \bibinfo {pages} {083506} (\bibinfo {year} {2024})},\ \Eprint {http://arxiv.org/abs/2303.07980} {arXiv:2303.07980 [astro-ph.CO]} \BibitemShut {NoStop}%
\bibitem [{\citenamefont {Young}\ \emph {et~al.}(2022)\citenamefont {Young}, \citenamefont {Musco},\ and\ \citenamefont {Byrnes}}]{Young:2022gfc}%
  \BibitemOpen
  \bibfield  {author} {\bibinfo {author} {\bibfnamefont {S.}~\bibnamefont {Young}}, \bibinfo {author} {\bibfnamefont {I.}~\bibnamefont {Musco}}, \ and\ \bibinfo {author} {\bibfnamefont {C.~T.}\ \bibnamefont {Byrnes}},\ }\href {\doibase 10.1088/1475-7516/2022/02/032} {\bibfield  {journal} {\bibinfo  {journal} {JCAP}\ }\textbf {\bibinfo {volume} {02}},\ \bibinfo {pages} {032} (\bibinfo {year} {2022})},\ \Eprint {http://arxiv.org/abs/2111.11456} {arXiv:2111.11456} \BibitemShut {NoStop}%
\end{thebibliography}%

\end{document}